\documentclass[reprint, NumberedRefs]{JASAnew}

\usepackage[sort&compress]{natbib}

\begin{document}
%% the square bracket argument will send term to running head in
%% preprint, or running foot in reprint style.

%\title[DAS for ocean applications]{Distributed acoustic sensing for ocean applications}
\title[DAS for ocean applications]{Overview of distributed acoustic sensing: Theory and ocean applications}

%% repeat as needed
\author{Angeliki Xenaki}
%\affiliation{Centre for Maritime Research and Experimentation, Science and Technology Organization -- NATO, La Spezia, 19126, Italy}
\affiliation{Institute of Applied and Computational Mathematics, Foundation for Research and Technology--Hellas, Heraklion, Crete, Greece}
%% for corresponding author
%\email{Angeliki.Xenaki@cmre.nato.int}
\email{Angeliki.Xenaki@iacm.forth.gr}
\author{Peter Gerstoft}
\affiliation{Noiselab, Scripps Institution of Oceanography, University of California San Diego, La Jolla, California 92093-0238, USA}
\author{Ethan Williams}
\affiliation{Department of Earth and Space Sciences,
University of Washington, Seattle, Washington 98195, USA}
\author{Shima Abadi}
\affiliation{School of Oceanography, University of Washington, Seattle, Washington 98195, USA}

\date{\today} 
\begin{abstract}
Extensive monitoring of acoustic activities is important for many fields, including biology, security, oceanography, and Earth science.
Distributed acoustic sensing (DAS) is an evolving technique for continuous, wide-coverage measurements of mechanical vibrations, which is suited to ocean applications.
DAS illuminates an optical fiber with laser pulses and measures the backscattered wave due to small random variations in the refractive index of the material. External stimuli, such as mechanical strain due to acoustic wavefields impinging on the fiber-optic cable, modulate the backscattered wave. Continuous measurement of the backscattered signal provides a distributed sensing modality of the impinging wavefield.
Considering the potential use of existing telecommunication fiber-optic cables deployed across the oceans, DAS has emerged as a promising technology for monitoring the underwater soundscape.
This review presents advances in DAS in the last decade and details the underlying physics from electromagnetic to mechanical and eventually acoustic quantities.
To guide the use of DAS for ocean applications, the effect of DAS acquisition parameters in signal processing is explained.
Finally, DAS is demonstrated on data from the Ocean Observatories Initiative Regional Cabled Array for the detection of sound sources, such as whales, ships, and earthquakes.
\end{abstract}

%% pacs numbers not used

\maketitle

%  End of title page for Preprint option --------------------------------- %

\section{\label{sec:Intro} Introduction}

Light propagating in an optical fiber undergoes backscattering due to small random variations in the refractive index of the material.
Fiber-optic sensing technology is based on the phase modulation of the backscattered light traveling in an optical fiber due to external stimuli such as mechanical strain or temperature changes.
Distributed sensing transforms a fiber cable into a very long, dense sensor array\cite{grattan2000fiber} that can extend over hundreds\cite{waagaard2021realtime} 
or even thousands of kilometers,\cite{mazur2024real, ronnekleiv2025range} offering unprecedented sensing coverage with rapidly advancing technology.
Moreover, fiber-optic sensing makes temperature and vibration measurements in harsh environments accessible due to its robustness to electromagnetic interference, high pressure, acidity, and temperature.\cite{Miah2017, Lindsey2021} 

Distributed acoustic sensing (DAS) is a fiber-optic sensing technology that illuminates an optical fiber with laser pulses and measures phase differences of the backscattered wave along the fiber.\cite{Hartog2017}
External mechanical vibrations will stress a fiber-optic cable that is in mechanical contact with the surrounding medium.
Strain, i.e., the deformation of the fiber under stress, alters the two-way distance traveled by the laser pulse, changing the phase of the backscattered light.\cite{KennettBook}
DAS records the differential phase of the backscattered light to estimate the strain (or strain rate) along the fiber, which is in turn related to the displacement (or velocity) of the mechanical vibration.\cite{masoudi2016contributed, yang2022sub}
Elongated optical fibers are mostly sensitive to the axial component of the induced strain.\cite{martin2021introduction, williams2024toward}
To overcome the effect of the directional sensitivity of DAS to the measured mechanical wavefield\cite{fang2023directional}, helically wound fiber-optic cables have been proposed.\cite{kuvshinov2016interaction, ning2018multicomponent}
Given that stress and strain are related through the elastic modulus of the solid material according to Hooke's law,\cite{Brekhovskikh1994} 
both the fiberglass coating\cite{budiansky1979pressure} and the deployment conditions of the cable, e.g., burial depth and seafloor type,\cite{sladen2019distributed} affect the strain sensitivity of the fiber-optic cable to the applied stress.

DAS has gained momentum in geophysics as a cost-effective, high-resolution method for seismic monitoring and seismic tomography.\cite{mateeva2014distributed, zhan2020distributed, Lindsey2021}
Applications range from near-surface seismic monitoring of ambient noise\cite{dou2017distributed, saengduean2024multi} to borehole measurements of hydraulic fracturing\cite{jin2017hydraulic, chien2023automatic} and monitoring the sea state dynamics with submarine cables.\cite{lindsey2019illuminating, lin2024measurement,chien2025calibration}
Recently, the potential of DAS technology has been leveraged for monitoring the underwater acoustic wavefield with promising results in the identification and localization of low-frequency sound sources such as blue and fin whale vocalizations\cite{bouffaut2022eavesdropping, landro2022sensing, wilcock2023distributed, Goestchel2025fin} and ship noise.\cite{rivet2021preliminary,shao2025tracking,paap2025leveraging}
DAS converts the length of the fiber cable into an extensive linear aperture, which is especially useful for localization of low-frequency sound sources (below 100 Hz).\cite{rivet2021preliminary, bouffaut2022eavesdropping, landro2022sensing, wilcock2023distributed}
Monitoring the ocean soundscape at low frequencies\cite{wilcock2014sounds} is key to understanding the impact of anthropogenic noise\cite{andrew2011long, chapman2011low, mcdonald2006increases} and climate change\cite{gavrilov2008long} on ecology,\cite{gavrilov2012steady} security, and marine exploration.\cite{miksis2016low}
So far, the characterization of the ocean soundscape has been based on recordings from hydrophones and hydrophone arrays with limited spatial coverage.\cite{mcdonald2006increases, chapman2011low}

Despite the increasing interest in DAS technology for high-resolution observations of acoustic signals, the current applications in ocean acoustics are limited to the aforementioned low-frequency wavefields, i.e., whale calls, ship noise, and ocean swell.
The absence of ocean acoustic observations at higher frequencies is related to the choice of acquisition parameters of the current DAS systems, such as the interrogated fiber length and the pulse and gauge length, which are set with the objective of sufficient coverage and signal-to-noise ratio, respectively.
Nevertheless, calibration experiments with controlled transmitted signals\cite{douglass2023distributed, shen2024high,rychen2023test} and at-sea experiments for underwater communications\cite{potter2024first} demonstrate that DAS detection capability is not limited to low-frequency sounds.
This paper links the acquisition parameters with signal processing and provides insight for improved observations of acoustic signals.

After trend removal and band-pass filtering to the frequency range of interest, the data are commonly analyzed with interferometric methods,\cite{dou2017distributed, williams2022surface, saengduean2024multi} $f$-$k$ decomposition, and delay-and-sum beamforming.\cite{nasholm2022array}
Given the high dimensionality of DAS data, machine learning provides compelling methods for dimensionality reduction and feature clustering.\cite{chien2023automatic}
 
In this paper, we provide an overview of DAS technology. 
We detail the mathematical derivation of the underlying physics, from electromagnetic to mechanical and eventually acoustic quantities in Section~\ref{sec:DAS}.
The mathematical analysis highlights the physical principles of DAS in a generic manner, aiming to inform the choice of data acquisition parameters rather than to detail the technological specificity of different systems, which are evolving rapidly.
Section~\ref{sec:Beampattern} compares the spatial sensitivity of DAS to that of an array of spatially distributed sensors of a sizeable aperture.
Section~\ref{sec:DataAnalysis} analyzes publicly available data for sound source detection based on the phase of the acoustic particle displacement.
Finally, Section~\ref{sec:Conclusions} concludes the paper.

\begin{table*}[t]
\renewcommand{\arraystretch}{1.3}
\caption{\label{tbl:ListOfParameters} List of parameters and operators.}
\centering
\begin{tabular}{ | p{0.7cm} | p{3.5cm} | p{0.7cm} | p{3.5cm} | p{0.7cm} | p{3.5cm} |p{0.8cm} | p{3.2cm} |}
\hline
\multicolumn{2}{|c|}{Space} & \multicolumn{2}{c|}{Time} & \multicolumn{2}{c|}{Direction}& \multicolumn{2}{c|}{Operators}\\
\hline
\hline
$z$ & axial distance & $t$ & time & $k_{z}$ & acoustic wavenumber along the fibre & $\widetilde{[\cdot]}$ & complex quantity\\
\hline
$L_{h}$ & channel distance - distance between spatial samples & $\tau$  & two-way travel time & $\theta$ & incident angle & $[\cdot]^{*}$ & complex conjugation\\
\hline
$L_{p}$ & light pulse length & $T_{p}$ & light pulse duration (width) & $\mathbf{e}$ & directional unit vector & $\widehat{[\cdot]}$ & estimated quantity\\
\hline
$L_{g}$ & gauge length & $\Delta\Phi$ & differential phase of backscattered light & & & $[\cdot]*[\cdot]$ & convolution \\
\hline
$L_{f}$ & interrogated fiber length & $\tau_{\text{max}}$ & two-way travel time along fiber length  &  & &  $[\cdot]_{z}$ & projection to the fiber's axis\\
\hline
$f_{s}$ & sampling frequency (spatial sampling) & $f_{r}$ & pulse repetition frequency (temporal sampling) & & & $[\cdot]_{p}$ & quantity of the transmitted light pulse\\
\hline
& & & & & & $[\cdot]_{b}$ & quantity of the total backscattered light wave\\
\hline
& & & & & & $[\cdot]_{a}$ & quantity of the acoustic field\\
\hline
\end{tabular}
\end{table*}

\begin{figure}[t]
\centering
\includegraphics[width=88mm]{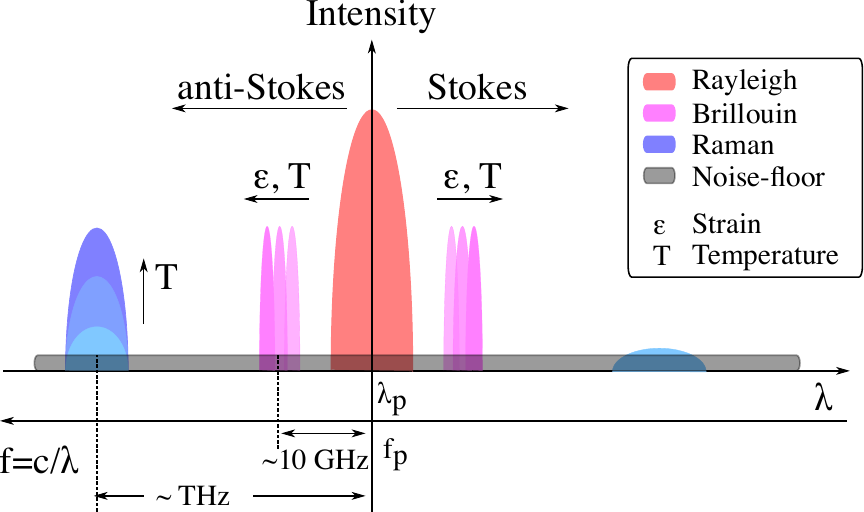}
\caption{(color online) Schematic representation of scattering mechanisms due to a light pulse of wavelength $\lambda_{p}$ and frequency $f_{p}$ propagating through an optical fiber. Single-mode optical fibers are typically driven with laser sources at wavelengths $\lambda_{p}$ of about 1550 nm ($f_{p} = c/\lambda_{p} \approx $  200 THz). External stimuli, such as an increase in temperature $T$ or strain $\epsilon$, affect the respective scattering response as indicated by the arrows.
The horizontal axis represents the wavelength $\lambda$ or, in reverse direction, the frequency $f$ of the backscattered light. Stokes processes refer to an up-shift of the wavelength of the scattered light $\lambda > \lambda_{p}$, whereas anti-Stokes processes refer to a down-shift $\lambda < \lambda_{p}$. The Stokes radiation for Raman scattering is relatively insensitive to temperature and has a lower intensity than the corresponding anti-Stokes radiation due to lower scattering efficiency at lower frequencies.\cite{Ukil2011}}
\label{fig:scattering_mechanisms}
\end{figure}

\section{\label{sec:DAS} Distributed Acoustic Sensing Technology}

DAS measures the effect of external mechanical vibrations on the backscattering of laser light traveling along an optical fiber.
This section details the DAS technology by providing an overview of the light scattering mechanisms within a fiber-optic cable (Sec. \ref{sec:ScatteringMechanisms}), deriving the mathematical expression for the backscattered wave under monochromatic excitation (Secs. \ref{sec:Backscatter}-\ref{sec:BackscatterFrequency}),  explaining the detection method of the backscattered light (Sec. \ref{sec:DiffCOTDR}), showing how the measurand is related to strain from external mechanical stress (Sec. \ref{sec:AxialStrain}), and finally explaining the typical acquisition parameters and the data structure (Sec. \ref{sec:DataFormat}).
Table~\ref{tbl:ListOfParameters} provides a list of the important parameters involved.

\subsection{\label{sec:ScatteringMechanisms} Scattering mechanisms}

An optical fiber is a thin thread of silica glass with a core of dielectric material and cladding with a lower refractive index. 
The difference in refractive index between the core and the cladding creates an optical waveguide for light transmitted in the fiber.
Optical fibers are designed to have a core with a constant refractive index such that light transmitted into the fiber travels undisturbed along the fiber.
However, in practice, small density variations in the core material due to impurities or the manufacturing procedure introduce small random variations in the refractive index along the optical fiber, which cause scattering of the propagating electromagnetic wave.\cite{KennettBook}
Hence, the refractive index along the fiber's axis can be described as
\begin{equation}
n(z)  = n_{f} + \mathrm{d} n(z),
\label{eq:AxilaProfileRefractiveIndex}
\end{equation}
where $n_{f}$ is the nominal refractive index of the optical fiber's core material and $\mathrm{d} n(z)$ describes the localized fluctuation of the refractive index due to material inhomogeneities.
The refractive index determines the speed of light in the optical fiber as $c_{n} \approx c/n_{f}$, where $c \approx 3 \times 10^8$~m/s is the speed of light in vacuum.

The main scattering mechanisms are Rayleigh, Brillouin, and Raman scattering.\cite{masoudi2016contributed}
Rayleigh scattering is an elastic process caused by local scatterers with dimensions much smaller than the wavelength that does not involve energy transfer to the scatterer and does not shift the frequency of the incident wave.
Brillouin and Raman scattering are inelastic processes caused by changes in the vibrational states in a lattice and molecular level, respectively, that involve energy transfer to the scatterer and, as a result, shift the frequency of the incident wave.\cite{Cotter1983, Shiota1992}
Brillouin scattering can be leveraged for both distributed strain and temperature measurements,\cite{Bao2011} whereas Raman scattering can be leveraged for distributed temperature sensing.\cite{Ukil2011}
Rayleigh scattering exhibits the highest intensity and is the main scattering mechanism exploited for DAS measurements.\cite{KennettBook}
Figure~\ref{fig:scattering_mechanisms} depicts the effect of external stimuli such as temperature and strain to elastic and inelastic optical scattering mechanisms [\onlinecite[Ch. 3.8]{Grattan2000}].
\begin{figure*}[t]
\centering
\includegraphics[width=145mm]{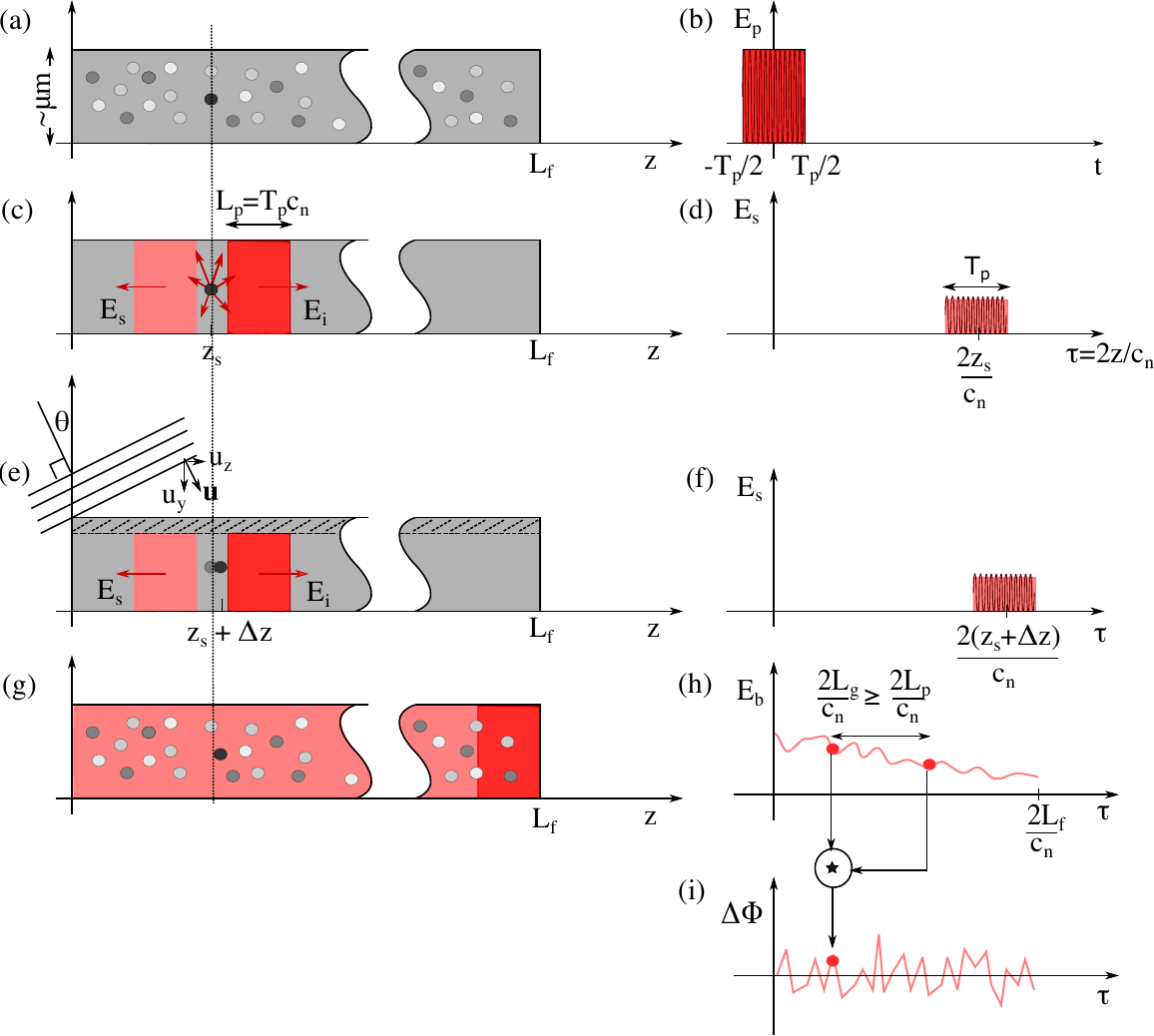}
\caption{(color online) Schematic representation of active optic sensing for passive monitoring of external vibrational fields.
(a) A single-mode fiber of total length $L_{f}$ and a cross-section in the order of $\mu$m with randomly distributed impurities in the core material.
(b) The fiber is interrogated with a laser pulse $E_{p}$ of duration $T_{p}$.
(c) As the probe pulse travels along the fiber at light speed $c_{n}$ extending over the pulse length $L_{p}$, a point scatterer located at a distance $z_{s}$ from the interrogator causes part of the incident energy $E_{i}$ to propagate back to the interrogator $E_{s}$.
(d) The backscattered wave $E_{s}$ from a single point scatterer is received around a time instant $\tau$ corresponding to the two-way travel time between the laser source and the scatterer with a duration equal to the probe pulse.
(e) An external vibration from an incident angle $\theta$ characterized by the spatial displacement vector $\mathbf{u}$ applies stress to the fiber, which deforms its shape. Under strain, the scatterers' positions are slightly perturbed.
(f) A scatterer's displacement $\Delta z$ introduces a proportional delay to the backscattered wave. 
(g) The total backscattered field results from the superposition of the backscattered light from all illuminated scatterers along the fiber.
(h) Cross-correlating instances of the total backscattered signal $E_{b}$ received from adjacent fiber locations separated by the gauge length $L_{g}$, provides (i) a differential phase measurement $\Delta\Phi = \angle{(E_{b}\star E_{b})(L_{g})}$, which is lineraly related to the axial strain along the fiber due to external vibrations; see Sec.~\ref{sec:AxialStrain}.
The fiber geometry is depicted as a function of distance $z$ from the interrogator at $z=0$ (left column), whereas quantities referring to the light waves are depicted as a function of time (right column), i.e., (b) one-way travel time $t=z/c_{n}$ and (d), (f), (h), (i) two-way travel time $\tau = 2z/c_{n}$, respectively.}
\label{fig:fiber}
\end{figure*}

\subsection{\label{sec:Backscatter} Rayleigh backscattered wave}

The active sensor in DAS technology is an optoelectronic system, referred to as the interrogation unit or, simply, the interrogator. The interrogator transmits a light pulse into the optical fiber and detects the intensity of the backscattered electric field due to the interaction of the transmitted pulse with the impurities along the fiber.\cite{Hartog2017}

The most common single-mode optical fibers are designed with a cross-section of micro-metric dimensions such that only a single mode (transverse mode) of electromagnetic wave propagation is supported without modal dispersion.
For a typical single-mode optical fiber, a minimum attenuation in the order of $0.2$~dB/km is achieved for a light wavelength of $1550$~nm [\onlinecite[Fig. 2.3]{Hartog2017}], i.e., at a frequency of $200$~THz.
Hence, a narrowband laser pulse at this frequency would sustain a minimum transmission loss.\cite{Miah2017, Hartog2017} 

Multimode optical fibers have a core with a larger cross-section (at least an order of magnitude larger than single-mode fibers), which guides multiple electromagnetic modes simultaneously [\onlinecite[Sec. 2.1]{Hartog2017}]. 
Modal dispersion of the transmitted light as it propagates along the fiber eliminates its phase coherence and polarization information.
Consequently, multimode optical fibers do not support highly sensitive detection methods, such as coherent optical reflectometry (see Appendix), resulting in reduced detection range and resolution.
However, light emitting diodes (LED) are used for transmission in multimode fibers instead of highly coherent laser sources required in single-mode fibers, resulting in cost-effective systems [\onlinecite[Ch. 1]{Grattan2000}].
Despite the increased cost, single-mode fibers and coherent detection systems are the standard configuration for DAS due to the higher sensitivity.

Figure~\ref{fig:fiber} depicts the mechanism of active optic sensing in single-mode fibers and indicates its utilization for passive sensing of external vibrations. 
The spatial dimension $z$ corresponds the axial distance from the interrogator, whereas the temporal dimensions $t=z/c_{n}$ and $\tau = 2z/c_{n}$ correspond to one-way (unidirectional) and two-way (bidirectional) travel time of a light wave along the fiber, respectively.

In mathematical terms, a probe pulse [Fig.~\ref{fig:fiber}(b)] of duration $T_{p}$ and phase $\phi(t)$, generated by the laser source at $z=0$ is expressed as
\begin{equation}
E_{p}(t) = p\left( \frac{t}{T_{p}}\right) \cos\left(\phi(t)\right).
\label{eq:ProbePulse}
\end{equation}
The pulse envelope $p$ is defined by the window function $w$ as\cite{harris2005use}
\begin{equation}
p\left( \frac{t}{T_{p}} \right) = 
\begin{cases}
w(t), \, & \text{for} \quad \lvert t \rvert \leq \frac{T_{p}}{2} \\
0, \, & \text{otherwise.}
\end{cases}
\label{eq:PulseLength}
\end{equation}
For example, for a rectangular window, $w(t)=1$.
For mathematical simplicity, the window is considered symmetric around the origin.
The pulse instantaneous phase is determined by the instantaneous angular frequency $\omega (t)$  as,
\begin{equation}
\phi(t) =\int_{-T_{p}/2}^{t} \omega(\mathfrak{t})\mathrm{d}\mathfrak{t}.
\label{eq:PulsePhase}
\end{equation}

\begin{figure}[t]
\centering
\includegraphics[width=88mm]{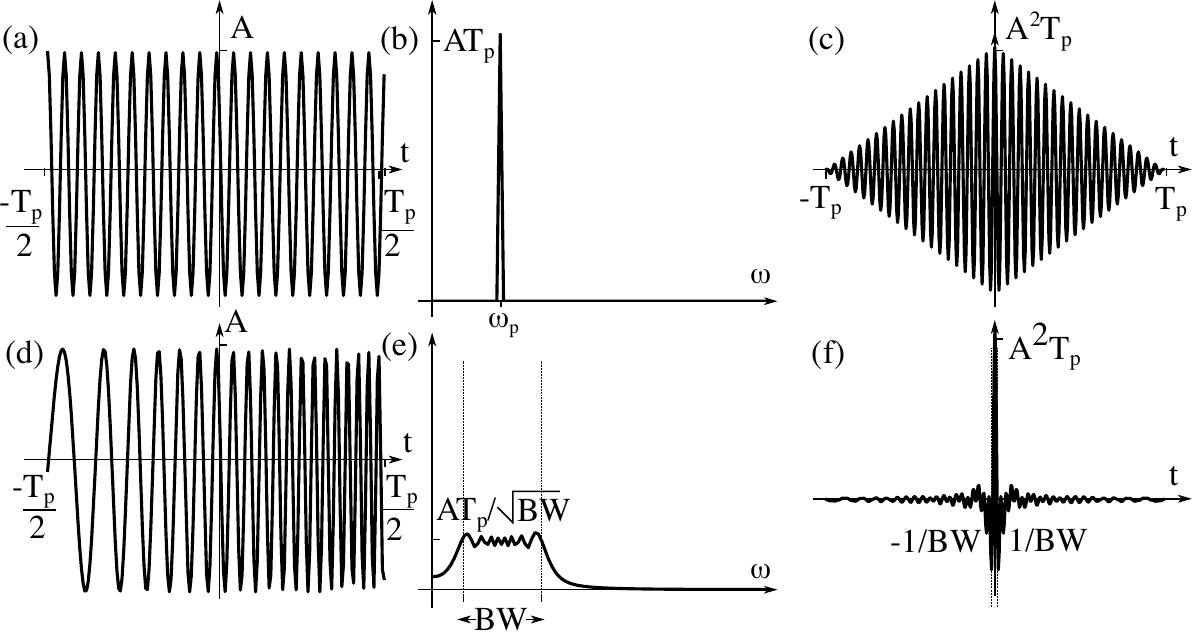}
\caption{Transmitted laser pulse characteristics. (a) Monochromatic pulse at angular frequency $\omega_{p}$ defined by a rectangular window of duration $T_{p}$, (b) its frequency spectrum, and (c) autocorrelation function. (d) Linear frequency modulated pulse of duration $T_{p}$ and bandwidth $BW$, (e) its frequency spectrum, and (f) autocorrelation function.} 
\label{fig:pulse_autocorrelation}
\end{figure}

For a monochromatic continuous wave (CW), the pulse has a constant angular frequency $\omega(t) = \omega_{p}$, such that, with an initial phase $\phi_{0}$,
\begin{equation}
\phi_{\text{CW}}(t) =\omega_{p} t+ \phi_{0}.
\label{eq:PulsePhaseCW}
\end{equation}
For a linearly frequency modulated chirped pulse (LFM), the angular frequency varies linearly with time within the pulse duration from
$\omega_{1}$ to $\omega_{2}$ with central frequency $\omega_{p} = (\omega_{1} + \omega_{2})/2$ and bandwidth $BW = \omega_{2}-\omega_{1}$ such that $\omega(t) = \omega_{p} + \left(BW/T_{p}\right)t$ and the pulse phase is,
\begin{equation}
\phi_{\text{LFM}}(t) = \left(\omega_{p} + \frac{BW}{2T_{p}}t\right)t+ \phi_{0}.
\label{eq:PulsePhaseLFM}
\end{equation}
Figure~\ref{fig:pulse_autocorrelation} illustrates the temporal, spectral, and autocorrelation characteristics of continuous wave and frequency modulated pulses.
Both continuous wave signals\cite{wilcock2023distributed} and frequency modulated chirps\cite{waagaard2021realtime} are used as probe pulses with coherent detection methods in DAS technology. 
In particular, the inherent frequency diversity in frequency modulated probe pulses results in a correlation length defined by the bandwidth rather than by the pulse length for $BW \cdot T_{p} > 1$.
Hence, broadband pulses offer robustness against optical noise but require electronics (photodetectors/digitizers) with higher bandwidth increasing the complexity and cost of the system.\cite{Ruiz2019distributed}

In the following, we adopt the monochromatic continuous wave model in Eq.~\eqref{eq:PulsePhaseCW} to simplify the mathematical derivations and to comply with the system used for the data analysis in later Sections.

Consider the impulse response received at the interrogator at $z = 0$ as a function of the arrival time $\tau$ due to an ideal point scatterer at $z_{s}$
as the two-way Green's function
\begin{equation}
g_{s}(\tau | z_{s}) = \left(g_{i} * g_{b} \right)(\tau) = \delta \left( \tau - 2\frac{z_{s}}{c_{n}}\right),
\label{eq:GreensFunctionPointScatterer}
\end{equation}
where $\delta$ denotes the Dirac delta function, $*$ is the convolution operator and $g_{i}$ and $g_{b}$ denote the Green's function for the incident wave at location $z_{s}$ and time $t$ due to an impulse source at $z_{0}=0$ and $t_{0}=0$ (forward propagation) and the backscattered wave received at $z_{0}=0$ at time $\tau = 2t$ due to an ideal scatterer at $z_{s}$ illuminated at time $t$ (backward propagation),
\begin{equation}
\begin{aligned}
& g_{i}(t, z_{s} | t_{0}, z_{0}) \!=\! \delta \!\left(t- t_{0} - \frac{z_{s}\! - z_{0}}{c_{n}}\!\right) \!=\! \delta\! \left( t - \frac{z_{s}}{c_{n}}\right),\\
& g_{b}(\tau, z_{0} | t, z_{s}) \!=\! \delta \!\left(\tau-t + \frac{ z_{0}\! - z_{s}}{c_{n}}\!\right) \!=\! \delta\! \left( t - \frac{z_{s}}{c_{n}}\right).
\end{aligned}
\label{eq:GreensFunctions}
\end{equation}
More generally, the backscattered wave [Fig.~\ref{fig:fiber}(d)] from a point scatterer at $z_{s}$ with scattering strength $s(z_{s})$ due to the excitation pulse in Eq.~\eqref{eq:ProbePulse} is
\begin{equation}
\begin{aligned}
E_{s}(\tau | z_{s}) &= E_{p}(\tau) * \left(g_{s}(\tau | z_{s})s(z_{s})\right) \\
&= E_{p} \left( \tau - 2\frac{z_{s}}{c_{n}}\right)s(z_{s}).
\end{aligned}
\label{eq:BackscatterPointScatterer}
\end{equation}

External vibrations, such as an impinging acoustic wavefield at an incident angle $\theta$ characterized by particle displacement $\mathbf{u}$, apply stress to the fiber, which deforms its shape [Fig.~\ref{fig:fiber}(e)].
Under strain, any scatterer's position is slightly perturbed by $\Delta z$, resulting in a proportional delay at the associated backscattered signal, while the scattering strength $s(z_s)$ remains the same [Fig.~\ref{fig:fiber}(f)]:
\begin{equation}
\begin{aligned}
E_{s}(\tau | z_{s} + \Delta z) &= E_{p}(\tau) * \left(g_{s}(\tau | z_{s} + \Delta z)s(z_{s})\right) \\
&= E_{p} \left( \tau - 2\frac{z_{s} + \Delta z}{c_{n}}\right)s(z_{s}).
\end{aligned}
\label{eq:BackscatterPointScattererPerturbed}
\end{equation}

The total backscattered field received at $\tau = 2z/c_{n}$  results from the superposition of the backscattered light from all scatterers within the illuminated fiber length at $t=z/c_{n}$ defined by the pulse length $L_{p} = T_{p}c_{n}$, since $p\left((t-t_{s})/T_{p}\right) = w(t-t_{s})$ for $\lvert t-t_{s}\rvert \leq T_{p}/2$ is equivalent to $p\left((z-z_{s})/L_{p}\right) = w(z-z_{s})$ for $\lvert z-z_{s}\rvert \leq L_{p}/2$ with a linear transformation of variables between time and space through the speed of light.
Making use of Eqs.~\eqref{eq:BackscatterPointScatterer}, \eqref{eq:ProbePulse},\eqref{eq:PulseLength} and \eqref{eq:PulsePhaseCW} yields
\begin{equation}
\begin{aligned}
E_{b}(\tau) &\!= \!\!\!\! \!\!  \int\limits_{z-L_{p}/2}^{z+L_{p}/2} \!\!\!\! \!\!  E_{s}(\tau | z_{s}) \mathrm{d}z_{s} \!\! \!\ =  \!\! \!\! \!\! \!\!\int\limits_{z-L_{p}/2}^{z+L_{p}/2} \!\!\!\! \!\!  E_{p} \left( \tau \!-\! 2\frac{z_{s}}{c_{n}}\right) s(z_{s}) \mathrm{d}z_{s}\\
&\!=\!\!\!\!\!\! \int\limits_{z-L_{p}/2}^{z+L_{p}/2} \!\!\!\!\!  w\!\left(\! z \!-\!z_{s}\! \right) \! \cos\!\left( \! \omega_{p} \! \left(\! \tau \!-\! 2\frac{z_{s}}{c_{n}}\right) \!\! + \!  \phi_{0}\!\right)
 \!\! s(z_{s}) \mathrm{d}z_{s}.
\end{aligned}
\label{eq:BackscatterTotal}
\end{equation}
Then, with wavenumber $k_{p}=\omega_{p} /c_{n}$, the received time signal of the total backscattered field is mapped to the spatial dimension along the fiber's axis as [Fig.~\ref{fig:fiber}(h)]:
\begin{equation}
\begin{aligned}
E_{b}(z)
&\! =\!\!\!\!\!\!\int\limits_{z-L_{p}/2}^{z+L_{p}/2} \!\! \!\!\!\! w(z-z_{s}) \cos\left( 2 k_{p} (z\!- \!z_{s}) \! + \! \phi_{0}\right) s(z_{s}) \mathrm{d}z_{s}\\
&=\left( E_{p} * s \right )(z).
\end{aligned}
\label{eq:BackscatterTotalSpace}
\end{equation}
The resulting backscattered field in Eq.~\eqref{eq:BackscatterTotalSpace} is a stochastic process resulting from the convolution of the pulse smoothing kernel with the scattering strength $s$, which is a random function of the spatial dimension along the fiber's axis $z$.

\subsection{\label{sec:BackscatterFrequency} Spectral analysis of the backscattered field}

An alternative formulation for the backscattered field results from complex analysis.
Signifying complex quantities with the accent $\widetilde{\cdot}$, the Fourier transform $\widetilde{X}(\omega)$ of a signal $x(\tau)$ is defined as [\onlinecite[Eq. 4.1.32]{Proakis1996}],
\begin{equation}
\widetilde{X}(\omega) = \int\limits_{-\infty}^{\infty} x(\tau) e^{-j\omega\tau} \mathrm{d}{\tau},
\label{eq:FourierTransform}
\end{equation}
and the spatial Fourier transform is obtained by replacing $\omega = k c_{n}$ and $\tau=2z/c_{n}$, 
\begin{equation}
\widetilde{X}(k) = \int\limits_{-\infty}^{\infty} x(z) e^{-j2kz} \mathrm{d}{z}.
\label{eq:SpatialFourierTransform}
\end{equation}

In complex notation, the probe pulse in Eq.~\eqref{eq:ProbePulse} for monochromatic transmission in Eq.~\eqref{eq:PulsePhaseCW} is expressed as
\begin{equation}
\widetilde{E}_{p}(t) = p\left( \frac{t}{T_{p}}\right) e^{j (\omega_{p} t + \phi_{0}) },
\label{eq:ProbePulseComplex}
\end{equation}
and in the frequency domain,
\begin{equation}
\begin{aligned}
\widetilde{E}_{p}(\omega) &= \int\limits_{-\infty}^{\infty}
p\left( \frac{t}{T_{p}}\right) e^{j (\omega_{p} t + \phi_{0}) }e^{-j\omega t} \mathrm{d}{t}\\
&= \int\limits_{-T_{p}/2}^{T_{p}/2}
w(t) e^{-j(\omega-\omega_{p}) t} e^{j \phi_{0} }\mathrm{d}{t}\\
&= \widetilde{W}(\omega-\omega_{p})e^{j \phi_{0} },
\end{aligned}
\label{eq:ProbePulseSpectrum}
\end{equation}
where $\widetilde{W}(\omega-\omega_{p})$ denotes the temporal Fourier transform $\widetilde{W}(\omega)$ of the window function $w(t)$ centered at the angular frequency of the pulse $\omega_{p}$.

Similarly, the total backscattered field in Eq.~\eqref{eq:BackscatterTotal} is expressed as 
\begin{equation}
\widetilde{E}_{b}(\tau) =e^{j \phi_{0}}\!\!\!\!\int\limits_{z-L_{p}/2}^{z+L_{p}/2} \!\!\!\!\! w\!\left( z \!-\!z_{s} \right) \! e^{j \omega_{p}  \left( \tau \!-\! 2\frac{z_{s}}{c_{n}}\right) }
 s(z_{s}) \mathrm{d}z_{s},
\label{eq:BackscatterTotalComplex}
\end{equation}
and in space from Eq.~\eqref{eq:BackscatterTotalSpace},
\begin{equation}
\widetilde{E}_{b}(z) = e^{j \phi_{0}} \!\!\!\! \int\limits_{z-L_{p}/2}^{z+L_{p}/2}\!\!\!\! w(z-z_{s})e^{j 2 k_{p}( z- z_{s}) } s(z_{s}) \mathrm{d}z_{s}.
\label{eq:BackscatterTotalSpaceComplex}
\end{equation}

With $\widetilde{W}(k)$ and $\widetilde{S}(k)$ denoting the spatial Fourier transform of the pulse window $w(z)$ and the scattering function $s(z)$, respectively, the spatial Fourier transform of the total backscattered field in Eq.~\eqref{eq:BackscatterTotalSpaceComplex} is,
\begin{equation}
\widetilde{E}_{b}(k)\!\!=\!\!e^{j\phi_{0}} \!\! \left[ \widetilde{W}(k\!-\!k_{p})\widetilde{S}(k)\right]\!\! =\!\! e^{j\phi_{0}}  \!\!\left[ \widetilde{W}(k)\widetilde{S}(k\!-\!k_{p})\right]\!.
\label{eq:BackscatterTotalSpaceSTFT}
\end{equation}
For example, considering the pulse window to be a rectangular function, $\widetilde{W}(k)$ is a sinc function, as detailed later in Eq.~\eqref{eq:PulseAperture}. 
The scattering function $s(z)$ is a random process with zero mean and autocorrelation function $R_{ss}$.
Assuming that the scatterers in the fiber are spatially uncorrelated with variance $\sigma^{2}$, then $R_{ss}(z) = \sigma^{2}\delta\!\left(z\right)$. 
Hence, the Fourier transform of the scattering function is a random process with constant spectral density as results from the Fourier transform of the autocorrelation function $\widetilde{C}_{s}(k) = \mathcal{F} (R_{ss}(z)) = \sigma^{2}$.

\subsection{\label{sec:DiffCOTDR} Differential phase-measuring coherent optical reflectometry}

DAS systems employ phase-sensitive coherent optical reflectometry methods (see Appendix) to measure the phase shift in the backscattered light from two fiber locations [Fig.~\ref{fig:fiber}(h)].
These differential phase measurements [Fig.~\ref{fig:fiber}(i)] are linearly related to strain from mechanical vibrations and are used as proxy to strain estimates [\onlinecite[Ch.\ 6]{Hartog2017}].

There are several implementations of differential phase-measuring methods.
The underlying principle is mixing the backscattered signal with itself at a relative delay and detecting the phase of the resulting cross-correlation product.
The two signals can be mixed either in the optical domain (direct detection techniques) or the electrical domain (coherent detection techniques). They can have the same frequency (interferometric recovery technique and homodyne detection) or slightly different frequencies (dual-pulse systems or heterodyne detection).\cite{Hartog2017, He2021}

In all cases, the resulting interferometric field is detected with coherent optical reflectometry methods, Eq.~\eqref{eq:CurrentPhotodetector}, by mixing the backscattered light $\widetilde{E}_{b}(\tau)$ with a reference signal, which is the corresponding backscattered light at a relative delay of $\Delta\tau$, $E_{\text{ref}} = \widetilde{E}_{b}(\tau + \Delta\tau)$.
For an undisturbed fiber, this delay corresponds to the constant gauge length $L_{g}$ along the fiber.
External mechanical vibrations deform the optical fiber cable and perturb the gauge length by $\Delta L$ resulting in a relative delay of $\Delta\tau = 2(L_{g} + \Delta L)/c_{n}$.
Hence, the interference product at the photodetector is [\onlinecite[Eq.\ (6.11)]{Hartog2017}]
\begin{equation}
\begin{aligned}
&\widetilde{E}_{b}\left( \tau \right)\widetilde{E}^{*}_{b}\left(\tau +\Delta\tau \right)
\\
&= \Big( e^{j \phi_{0}}\!\!\!\!\! \int\limits_{z-L_{p}/2}^{z+L_{p}/2} \!\!\!\!\!  w\left( z -z_{s} \right)  e^{j \omega_{p}  \left( \tau - 2\frac{z_{s}}{c_{n}}\right) }
 s(z_{s}) \mathrm{d}z_{s}  \Big) 
\\
&\cdot\Big(e^{j \phi_{0}}\!\!\!\!\!\!\!\!\! \!\!\!\!\!  \int\limits_{z + L_{g} + \Delta L -L_{p}/2}^{z + L_{g} + \Delta L +L_{p}/2} \!\!\!\!\!\!\!\!\! \!\!\!\!\!   w\left( z \!+\! L_{g} + \Delta L  \!-\! z_{s} \right) e^{j \omega_{p}  \left( \tau + \Delta\tau - 2\frac{ z_{s}}{c_{n}}\right) }
 s(z_{s}) \mathrm{d}z_{s}\Big)^*,
\end{aligned}
\label{eq:DifferentialPhaseMeasurements}
\end{equation}
or expressed in the spatial domain, 
\begin{equation}
\begin{aligned}
&\widetilde{E}_{b}\left( z \right)\widetilde{E}^{*}_{b}\left(z + L_{g} + \Delta L \right)\\
& = \! e^{j (2k_{p}z +\phi_{0})} \!\!\!\!\! \int\limits_{z-L_{p}/2}^{z+L_{p}/2} \!\!\!\! w(z\!-\!z_{s}) e^{-j 2 k_{p} z_{s}}s(z_{s}) \mathrm{d}z_{s}
\\
& \cdot \! e^{-j(2k_{p}(z + L_{g} + \Delta L) + \phi_{0})} \Big(\!\!\!\!\! \!\!\!\!\! \!\!\!\!\! \int\limits_{z + L_{g} + \Delta L-L_{p}/2}^{z+ L_{g} + \Delta L + L_{p}/2} \!\!\!\!\!\!\!\!\! \!\!\!\!\! \!\!\!  w(z \!-\! z_{s}) e^{-j 2 k_{p} z_{s} } s(z_{s})\mathrm{d}z_{s}\!\Big)^*,
\end{aligned}
\label{eq:DifferentialPhaseMeasurementsSpace}
\end{equation}
with wavenumber spectrum,
\begin{equation}
\begin{aligned}
\widetilde{E}_{b}\left( k \right)\widetilde{E}^{*}_{b}(k)& 
= e^{-j2 k_{p} (L_g+\Delta L)} \\& \quad \cdot \left[ \widetilde{W}(k)\widetilde{S}(k\!-\!k_{p})\!\right]
\!\!\left[ \widetilde{W}(k)\widetilde{S}(k\!-\!k_{p})\!\right]^{*}
\\
& = e^{-j2 k_{p} (L_g+\Delta L)} |\widetilde{W}(k)|^2 |\widetilde{S}(k-k_{p})|^2
\\
& = e^{-j\Delta\Phi} \widetilde{C}_{b(L_{g})}(k),
\end{aligned}
\label{eq:DifferentialPhaseMeasurementsWavenumber}
\end{equation}
where $\Delta\Phi  = 2k_{p} \Delta L$ is the differential phase and $\widetilde{C}_{b(L_{g})}(k) = 
|\widetilde{W}(k)|^2 |\widetilde{S}(k-k_{p})|^2e^{-j2k_p L_g}$ 
denotes the cross-spectral density of the backscattered signal at a relative delay determined by the gauge length $L_{g}$.

The gauge length determines the amount of overlap of the backscattered signals in Eq.~\eqref{eq:DifferentialPhaseMeasurements}, i.e., $L_{\text{overlap}} = \max\left(L_{p} - L_{g}, 0\right)$.
The mixed signals are correlated for a gauge length shorter than the pulse length, $L_{g} < L_{p}$.
Hence, the gauge length should be longer than the pulse length to measure the effect of phase modulation due to external vibrations independently of the pulse that probes the fiber. 
\begin{figure}[t]
\centering
\includegraphics[width=88mm]{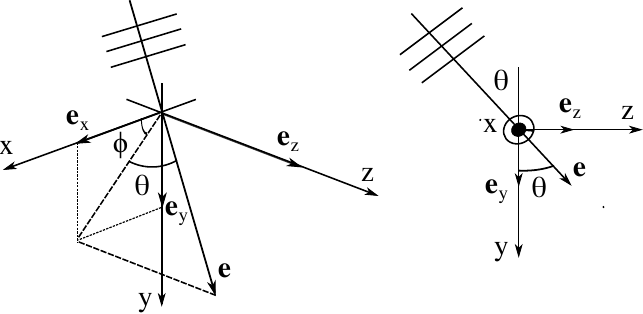}
\caption{Three-dimensional Cartesian coordinate system indicating the direction $\mathbf{e}$ of an impinging wavefield in terms of the angle $\phi$ defined anti-clockwise between the $x$ and the $y$-axis and the angle $\theta$ defined anti-clockwise between the $xy$-plane and the $z$-axis. By convention, the linear optical fiber cable is aligned with the z-axis. The two-dimensional perspective shows that the angle $\theta$ corresponds to the wavefield's angle of incidence.} 
\label{fig:spherical2cartesian}
\end{figure}
\begin{figure*}[t]
\centering
\includegraphics[width=110mm]{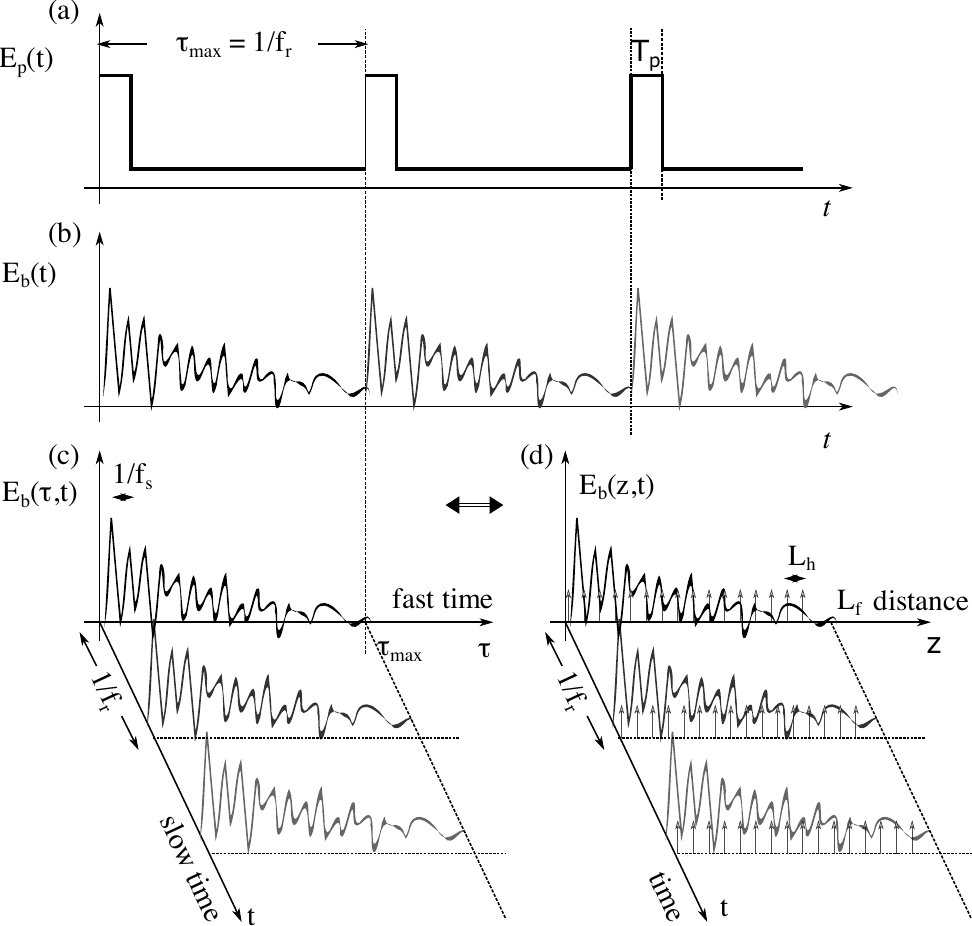}
\caption{Two-dimensional data structure in DAS [adapted from \onlinecite[Fig. 6.1]{Hartog2017}]. 
(a) An interrogator uses a laser source to transmit a pulse train $E_{p}(t)$ into the optical fiber, and (b) receives the backscattered signal $E_{b}(t)$ due to inhomogeneities in the dielectric material.
(c) Back at the laser's position, the received time series is converted to slow time $t$ and fast time $\tau$, which corresponds to the two-way travel time of the backscattered wave along the fiber. The time scales are determined by the acquisition sampling frequency $f_{s}$ along the fast-time axis $\tau$, and the pulse repetition frequency $f_{r}$ along the slow-time axis $t$. 
(d) Measurements along the fast-time correspond to the spatial dimension along the fiber, $z=\tau c_{n}/2$, and are commonly downsampled to the channel spacing $L_{h}$.
Phase variations in the backscattered signal measured along distance $z$ from the interoggator are associated with the applied axial strain $\epsilon_{zz}$, whereas phase variations in the backscattered signal measured across time $t$ are associated with the strain rate $\partial\epsilon_{zz}/\partial t$.}
\label{fig:slow_fast_time}
\end{figure*}

\subsection{\label{sec:AxialStrain} Axial strain estimates}

The following analysis shows that the differential phase measurements of the light backscatter (optics) are linearly related to the axial fiber strain (mechanics), which, in turn, is non-linearly related to acoustic quantities in the environment (acoustics).
The subscript $a$ is introduced to indicate acoustic quantities and differentiate them from the optical quantities in the previous analysis.
Namely,  $\omega_{a}$ denotes the angular frequency of the acoustic wave,  $k_{a} =\omega_{a}/c_{a}$ denotes the wavenumber, and $c_{a}$ denotes the sound speed.

Interrogating the optical fiber cable with phase-sensitive coherent reflectometry methods provides a distributed measurement of the differential phase $\Delta\Phi$ caused by optical path length variations $\Delta L$; see Eq.~\eqref{eq:DifferentialPhaseMeasurementsWavenumber}.
Specifically,
\begin{equation}
\begin{aligned}
\Delta\Phi  &= 2 k_{p} \Delta L = 2\frac{2\pi (n+\Delta n)}{\lambda_{p}}L_{g}\frac{\Delta L}{L_{g}}\\
&=
\frac{4\pi n\xi L_{g}}{\lambda_{p}}\frac{\Delta L}{L_{g}},
\end{aligned}
\label{eq:DifferentialPhasevsStrain}
\end{equation}
where $\lambda_{p}$ is the wavelength of the probe pulse and $\xi = 1+\Delta n/n$ is the Pockels coefficient that accounts for the strain-optical effect, i.e., strain applied to the fiber both changes its length and modifies its refractive index.\cite{Hartog2017, williams2024toward}
In single-mode fiberglass, the Pockels coefficient is $\xi = 0.79$.\cite{Lindsey2021} 
Note that $\Delta n$ describes a change in the refractive index due to strain, not the differential change $\mathrm{d} n$ due to material inhomogeneities in Eq.~\eqref{eq:AxilaProfileRefractiveIndex}.

Considering an optical fiber cable that is coupled with an elastic medium, e.g., deployed on the seafloor, the cable's deformation $\Delta L$ is attributed to mechanical waves impinging on it [Fig.~\ref{fig:fiber}(e)].
Specifically, the particle displacement vector in three-dimensional space $\mathbf{r} = [x, y, z]^{T}$, due to an acoustic plane wave with amplitude $A$ is expressed as,
\begin{equation}
\widetilde{\mathbf{u}}(t, \mathbf{r}) = (A\mathbf{e})\odot e^{j(\omega_{a} t - k_{a}\mathbf{e}\odot\mathbf{r})},
\end{equation}
where the direction of the plane wave is defined by the unit vector $\mathbf{e}$ and the $\odot$ operator denotes element-wise multiplication.

The directional unit vector in Cartesian coordinates is defined as $\mathbf{e} =[\cos\phi\cos\theta, \sin\phi\cos\theta, \sin\theta]^{T}$, where $\phi$ is the anti-clockwise angle between the $x$ and $y$-axis and $\theta$ is the anti-clockwise angle between the $xy$-plane and the $z$-axis, as depicted in Fig.~\ref{fig:spherical2cartesian}.
By convention, the linear optical fiber cable is aligned with the $z$-axis, such that the angle $\theta$ corresponds to the incident angle of the acoustic plane wave. 
Alternatively, the direction of the plane wave can be defined in terms of the grazing angle $\theta_{g} = 90^{\circ} - \theta$, i.e., the complementary of the incident angle, such that $\mathbf{e} =[\cos\phi\sin\theta_{g}, \sin\phi\sin\theta_{g}, \cos\theta_{g}]^{T}$.

The particle displacement along the $z$ direction, $\widetilde{u}_{z}(t, z) = A\sin\theta e^{j(\omega_{a} t - k_{a}z\sin\theta)}$,  is related to the axial strain as [\onlinecite[Eq.\ (1-2)]{Ewing1957}]
\begin{equation}
\begin{aligned}
\widetilde{\epsilon}_{zz} &= \frac{\partial \widetilde{u}_{z}}{\partial z} = \underset{\ell \rightarrow 0}{\lim} \frac{\widetilde{u}_{z}(t, z + \frac{\ell}{2}) - \widetilde{u}_{z}(t, z - \frac{\ell}{2})}{\ell}\\
&\approx \frac{\Delta L}{L_{g}} = \frac{\widetilde{u}_{z}(t, z + \frac{L_{g}}{2}) - \widetilde{u}_{z}(t, z - \frac{L_{g}}{2})}{L_{g}}  \\
& = \widetilde{u}_{z}(t, z) \frac{e^{-jk_{a}\frac{L_{g}}{2}\sin\theta } - e^{jk_{a}\frac{L_{g}}{2}\sin\theta} }{L_{g}}\\
& = \widetilde{u}_{z}(t, z)\frac{-j2\sin(k_{a}\frac{L_{g}}{2}\sin\theta)}{L_{g}}\\
& = \widetilde{u}_{z}(t, z) \left(-j k_{a}\sin\theta\right)\frac{\sin(k_{a}\frac{L_{g}}{2}\sin\theta)}{k_{a} \frac{L_{g}}{2}\sin\theta}\\
& = \!\!-jk_{a}\sin^{2}\!\theta A e^{j(\omega_{a} t - k_{a}z\sin\!\theta )}\mathrm{sinc}(k_{z}L_{g}/2),
\end{aligned}
\label{eq:NormalStrain}
\end{equation}
where $k_{z} = k_{a}\sin\theta$ and the spatial derivative is approximated by the total deformation over a finite gauge length, ${\Delta L}/{L_{g}}$.

Equation~\eqref{eq:DifferentialPhasevsStrain} indicates that the measured differential phase is linearly related to an estimate of the axial strain over the gauge length $\widehat{\epsilon}_{zz} = \Delta L/L_{g}$ \eqref{eq:NormalStrain} as
\begin{equation}
\widehat{\epsilon}_{zz}= \frac{\lambda_{p}}{4\pi n \xi L_{g}}\Delta \Phi.
\label{eq:StrainMeasurement}
\end{equation}
In turn, the axial strain estimate in Eq.~\eqref{eq:StrainMeasurement} is related to the particle displacement of external vibrations such as the acoustic wavefield around the optical-fiber cable through Eq.~\eqref{eq:NormalStrain}, which reveals two important aspects for interpreting DAS measurements.

First, the dependence of the axial strain amplitude to the square of the sine of the wavefield's incident angle makes the strain measurement insensitive to broadside incidence, i.e., $\widetilde{\epsilon}_{zz} = 0$ for $\theta = 0$.
This angular dependency is crucial in interpreting DAS data, as it influences the amplitude and detectability of acoustic signals. It also limits the effectiveness of DAS in configurations where ray paths are predominantly orthogonal to the fiber-optic cable. To overcome this limitation, helically wound cables were developed offering broadside sensitivity and enabling the observation of signals from a wider range of angles.\cite{kuvshinov2016interaction, hornman2017field, al2023experiences}
Nevertheless, DAS is widely applied on fiber optic cables deployed for telecommunications, which comprise straight optical fibers.

Second, the finite difference approximation of the spatial derivative of the displacement over the gauge length introduces directivity to the measurement defined by the term $\mathrm{sinc}(k_{z}L_{g}/2)$. 
Hence, the gauge length acts as a spatial window which determines the measurement resolution; see Sec.~\ref{sec:Beampattern} for details.  

Some DAS systems, instead of estimating the axial strain in Eq.~\eqref{eq:StrainMeasurement} by unwrapping the differential phase measurement $\Delta\Phi$, estimate the strain rate by differentiating the phase measurement between successive pulses [\onlinecite[Sec.\ 6.3.2]{Hartog2017}], $\frac{\partial \epsilon_{zz}}{\partial t}$.
In such systems, denoting the pulse repetition period as $T$, the axial strain rate estimate is
\begin{equation}
\begin{aligned}
\frac{\partial \widetilde{\epsilon}_{zz}}{\partial t} &= \frac{\partial}{\partial t}\frac{\partial \widetilde{u}_{z}}{\partial z}\\
&\approx (-jk_{a}\sin\theta)\mathrm{sinc}(k_{z}L_{g}/2) \frac{\partial \widetilde{u}_{z}}{\partial t}\\
&\approx \omega_{a} k_{a}\sin^{2}\theta A e^{j(\omega_{a} t - k_{a}z\sin\theta)}\mathrm{sinc}(k_{z}L_{g}/2).
\end{aligned}
\label{eq:NormalStrainRate}
\end{equation}
Concisely, the axial strain is related to the particle displacement component along the fiber, whereas the axial strain rate is related to the corresponding component of particle velocity.

\subsection{\label{sec:DataFormat}  Data acquisition}

The preceding analysis indicates that setting up a DAS system requires defining a few acquisition parameters, such as channel spacing, pulse repetition rate, pulse width or pulse length, and gauge length.

The optical fiber of total length $L_{f}$ is interrogated at regular discrete spatial locations determined by the sampling frequency $f_{s}$.
After phase unwrapping, the measurements are downsampled to fewer spatial locations referred to as channels.
The channel distance $L_{h}$ is the distance between adjacent channels.

For unambiguous measurements, a new pulse is emitted only after the previous pulse has traveled along the fiber and returned to the transmission point.
Hence, the pulse repetition frequency in a DAS acquisition system depends on the length of the fiber $f_{r} = 1/\tau_{\text{max}} = c_n/(2L_{f})$.  
For example, for a 100~km optical fiber cable with a typical refractive index of $n \approx 1.46$, $f_{r}$ cannot exceed 1~kHz.

Figure~\ref{fig:slow_fast_time} illustrates the timescales of a typical DAS dataset as defined by the sampling frequency $f_{s}$ and the pulse repetition frequency $f_{r}$ [\onlinecite[Fig. 6.1]{Hartog2017}].
The output of a DAS system is structured in a two-dimensional format $E_{b}(\tau, t)$, where $\tau$ is the fast-time and $t$ is the slow-time.
Measurements along the fast-time correspond to the spatial dimension along the fiber, $z=\tau c_{n}/2$, and are commonly downsampled to the channel spacing $L_{h}$.

Measurements along the slow-time correspond to the temporal dimension of the dataset [Fig.~\ref{fig:slow_fast_time}].
The pulse repetition frequency defines the maximum observable acoustic frequency, $f_{a}^{\text{max}}<=f_{r}/2$.
The two-dimensional data arrangement resembles the typical data arrangement in synthetic aperture systems for radar [\onlinecite[Ch. 4.6.1]{sar2005}] or sonar applications.\cite{sas2011}
In active sensing with synthetic aperture systems, though, time and space are accessed by the probe pulse and the platform motion, respectively, hence, fast-time corresponds to the temporal and slow-time to the spatial dimension.

The laser pulse duration $T_{p}$, or pulse width, determines the pulse length $L_{p} = T_{p}c_{n}$.
The longer the pulse duration, the higher the SNR, but the lower the spatial resolution, as the fiber section within a pulse length acts like a single sensor.
A pulse length is typically longer than the channel distance, resulting in dependent channel measurements within the pulse length.
Typical pulse widths are 10--1000~ns, corresponding to a pulse length $L_{p} \approx$ 2--200~m for a fiber with a typical refractive index of $n \approx 1.46$.

The gauge length $L_{g}$ determines the overlap between the backscattered signals from adjacent locations along the fiber, which are mixed at the interrogator for differential phase measurements. 
Consequently, the gauge length should be larger than the pulse length for the mixed backscattered signals to be independent of the pulse coherence.
Additionally, the gauge length determines the resolution of the axial strain measurement along the fiber; see Eq.~\eqref{eq:NormalStrain}.
The greater the gauge length compared to the pulse length, the more linear the relation between differential phase and strain [\onlinecite[Fig. 6.29]{Hartog2017}].
On the other hand, the impinging wavefield should remain spatially coherent within a gauge length for meaningful measurements.
To fulfill these competing requirements, the gauge length is commonly set equal to the pulse length.\cite{spica2023pubdas,dataOOI}
\begin{figure}[t]
\centering
\includegraphics[width=88mm]{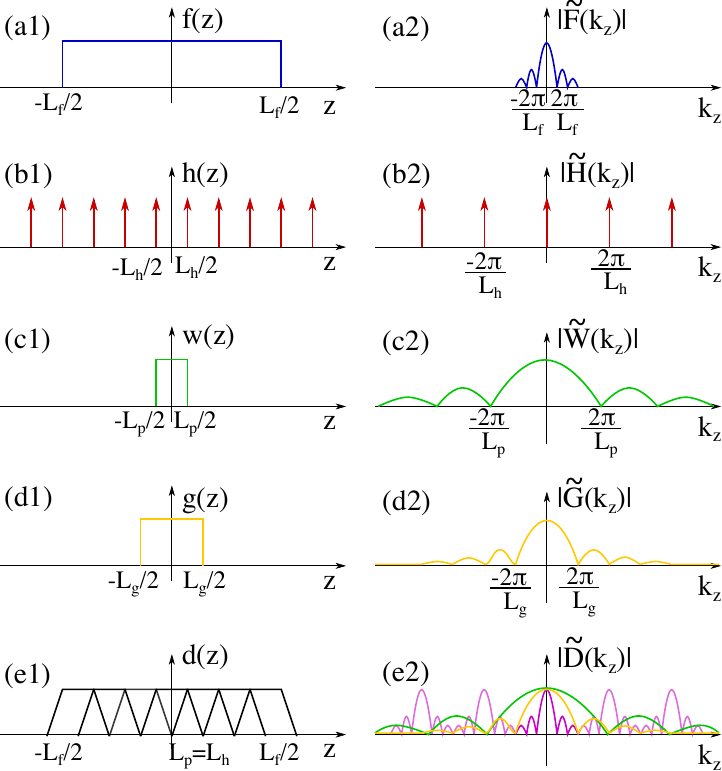}
\caption{(color online) Fourier pairs of distributed sensing parameters: Space (left) and wavenumber (right). (a) Fiber aperture with a total length of $L_{f}$, (b) spatial sampling at channel spacing $L_{h}$, (c) pulse aperture of length $L_{p}$, (d) gauge aperture of length $L_{g}$, and (e) total distributed sensing function resulting from sampling the total fiber aperture at the channel spacing (multiplication) and averaging over the pulse and gauge length (convolution). For unambiguous measurements, $L_{f}\gg L_{g}\geq L_{p}\geq L_{h}$.}
\label{fig:array_sampling}
\end{figure}
\begin{figure}[t]
\centering
\includegraphics[width=88mm]{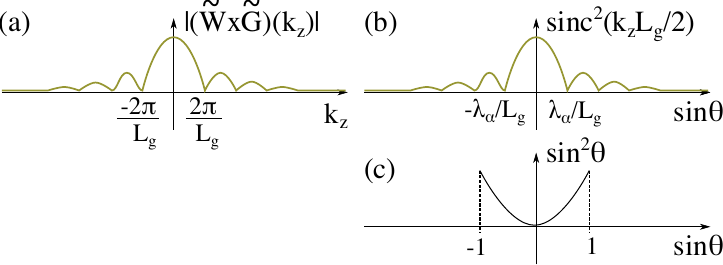}
\caption{(color online) DAS directivity when gauge length is equal to pulse length, $L_{g}$=$ L_{p}$, versus (a) the wavenumber along the fiber axis $k_{z}$ and (b) the direction of the incident angle $\sin\theta$. (c) The directional sensitivity of the axial strain.}
\label{fig:array_sampling_sintheta}
\end{figure}

\section{\label{sec:Beampattern} Distributed vs array processing}

Sensing wavefields through strain variations along the length of an optical fiber offers a distributed measurement modality, which samples densely large apertures.
Large apertures are required to accurately resolve the direction of arrival of the impinging wavefield, especially at low frequencies, which signifies the importance of distributed sensing compared to apertures of a limited extent that are discretely sampled by sensor arrays.\cite{VanTrees2002}

Figure~\ref{fig:array_sampling} illustrates the aperture sampling and the directivity of the distributed sensing modality. 
The optical fiber with length $L_{f}$ is represented as a rectangular function $\mathit{f}(z)$ in one-dimensional space $z$ and its Fourier transform $\widetilde{F}(k_{z})$ is the sinc function in the wavenumber domain $k_{z} = k_{a}\sin\theta$; Fig.~\ref{fig:array_sampling}(a1)--(a2),
\begin{equation}
\begin{aligned}
& \mathit{f}(z) = 
%\Pi (\frac{z}{L_{f}}) = 
\begin{cases}
1, \, & \text{for} \quad \lvert z \rvert \leq \frac{L_{f}}{2} \\
0, \, & \text{otherwise},
\end{cases}\\
& \widetilde{F}(k_{z}) = \mathcal{F}(\mathit{f}(z)) = \mathrm{sinc}(k_{z}L_{f}/2) = \frac{\sin (k_{z}L_{f}/2)}{k_{z}L_{f}/2}.
\end{aligned}
\label{eq:FiberAperture}
\end{equation}

The differential phase of the backscattered light from a laser pulse traveling along the optical fiber is sampled at $n$ channels distributed at regular spacings $L_{h}$ along the fiber, such that $L_{f} = n L_{h}$.
This sampling is represented by a Dirac comb function $h(z)$ with period $L_{h}$ and its Fourier transform is another Dirac comb function $\widetilde{H}(k_{z})$ with period $2\pi/L_{h}$; Fig.~\ref{fig:array_sampling}(b1)--(b2),
\begin{equation}
\begin{aligned}
& h(z) 
%=  \mathrm{III}_{L_{h}} (z) 
= \sum\limits_{i =-\infty}^{\infty} \delta(z- iL_{h}), \\
& \widetilde{H}(k_{z}) = \mathcal{F}(h(z)) 
%= \mathrm{III}_{1/L_{h}} (k_{z}) 
= \sum\limits_{i=-\infty}^{\infty} \delta(k_{z}- i\frac{2\pi}{L_{h}}),
\end{aligned}
\label{eq:ChannelSampling}
\end{equation}
where $\delta$ is the Dirac delta function and the sum is over all integers $i\in \mathbb{Z}$.

The fiber is probed by a narrowband light pulse defined by the window in Eq.~\eqref{eq:PulseLength}, which is represented here for illustration simplicity by a rectangular function $\mathit{w}(z)$ with Fourier transform $\widetilde{W}(k_{z})$; Fig.~\ref{fig:array_sampling}(c1)--(c2),
\begin{equation}
\begin{aligned}
& \mathit{w}(z) 
%= \Pi (\frac{z}{L_{p}}) 
= \begin{cases}
1, \, & \text{for} \quad \lvert z \rvert \leq \frac{L_{p}}{2} \\
0, \, & \text{otherwise},
\end{cases}\\
& \widetilde{W}(k_{z}) = \mathcal{F}(\mathit{w}(z)) = \mathrm{sinc}(k_{z}L_{p}/2) = \frac{\sin (k_{z}L_{p}/2)}{ k_{z}L_{p}/2}.
\end{aligned}
\label{eq:PulseAperture}
\end{equation}

Estimating strain over a finite gauge length introduces another spatial filter to the measurement of external vibrations; see Eq.~\eqref{eq:NormalStrain}. 
The gauge length can be represented as a rectangular aperture $\mathit{g}(z)$ with Fourier transform $\widetilde{G}(k_{z})$; Fig.~\ref{fig:array_sampling}(d1)--(d2),
\begin{equation}
\begin{aligned}
& \mathit{g}(z) 
%= \Pi (\frac{z}{L_{g}}) 
= \begin{cases}
1, \, & \text{for} \quad \lvert z \rvert \leq \frac{L_{g}}{2} \\
0, \, & \text{otherwise},
\end{cases}\\
& \widetilde{G}(k_{z}) = \mathcal{F}(\mathit{g}(z)) = \mathrm{sinc}(k_{z}L_{g}/2) = \frac{\sin (k_{z}L_{g}/2)}{k_{z}L_{g}/2}.
\end{aligned}
\label{eq:GaugeAperture}
\end{equation}

The resulting distributed aperture $d(z)$ is constructed from sampling the optical fiber at channel spacing, $(f \times h)(z)$, and convolving the result with the pulse. Due to averaging over the gauge length, the output is convolved with the gauge length function in Eq.~\eqref{eq:GaugeAperture}. 
Combining Eqs.~\eqref{eq:FiberAperture}--\eqref{eq:GaugeAperture} in either spatial or wavenumber domain yields
\begin{equation}
\begin{aligned}
&d(z) = ((f\times h)*w*g)(z),\\
&\widetilde{D}(k_{z}) = ((\widetilde{F}*\widetilde{H})\times \widetilde{W}\times \widetilde{G})(k_{z}),
\end{aligned}
\label{eq:TotalBeampattern}
\end{equation}
as depicted in Fig.~\ref{fig:array_sampling}(e1)--(e2).

The pulse length potentially spans multiple channels, $L_{p} = m L_{h}$ with $m \geq 1$.
Using a pulse length at least equal to the spatial sampling distance determined by the channel spacing results in a densely populated sensing array, i.e., an array without gaps between the sensors.
Such dense sensing attenuates effectively the grating lobes introduced by the spatial sampling; see Fig.~\ref{fig:array_sampling}(e2).

Since differential phase measurements are unambiguously related to strain from external vibrations only for $L_{g}\geq L_{p}$ (see Sec.~\ref{sec:DiffCOTDR}), it is the gauge length that determines the final spatial resolution for DAS measurements. Setting $L_{g} = L_{p}$ results in a directivity that is determined by,
\begin{equation}
\begin{aligned}
(\widetilde{W}\times \widetilde{G})(k_{z}) &= \mathrm{sinc}(k_{z}L_{p}/2)\mathrm{sinc}(k_{z}L_{g}/2) \\
&= \mathrm{sinc}^{2}(k_{z}L_{g}/2),
\end{aligned}
\label{eq:BeampatternVsSensitivity}
\end{equation}
as depicted in Fig.~\ref{fig:array_sampling_sintheta}(a) or, equivalently, in Fig.~\ref{fig:array_sampling_sintheta}(b) rescaling the wavenumber axis by $\lambda_{a}/(2\pi)$ to indicate the directional dependence $\sin\theta$.
The higher the frequency (smaller wavelengths) or the longer the gauge length, the more directive the DAS response to external vibrations becomes.
However, the axial strain estimate has a directional sensitivity with respect to the incident angle $\theta$, which is proportional  $\sin^2\theta$; see  Eq.~\eqref{eq:NormalStrain} and Fig.~\ref{fig:array_sampling_sintheta}(c).
Therefore, the more directive the DAS response is, the less sensitive the measurement is to external fields due to the directional response of the axial strain [Fig. \ref{fig:array_sampling_sintheta}].
Low frequencies with $\lambda_{a} \geq L_{g}$ are detected with higher sensitivity.
\begin{figure*}[t]
\centering
\includegraphics[width=150mm]{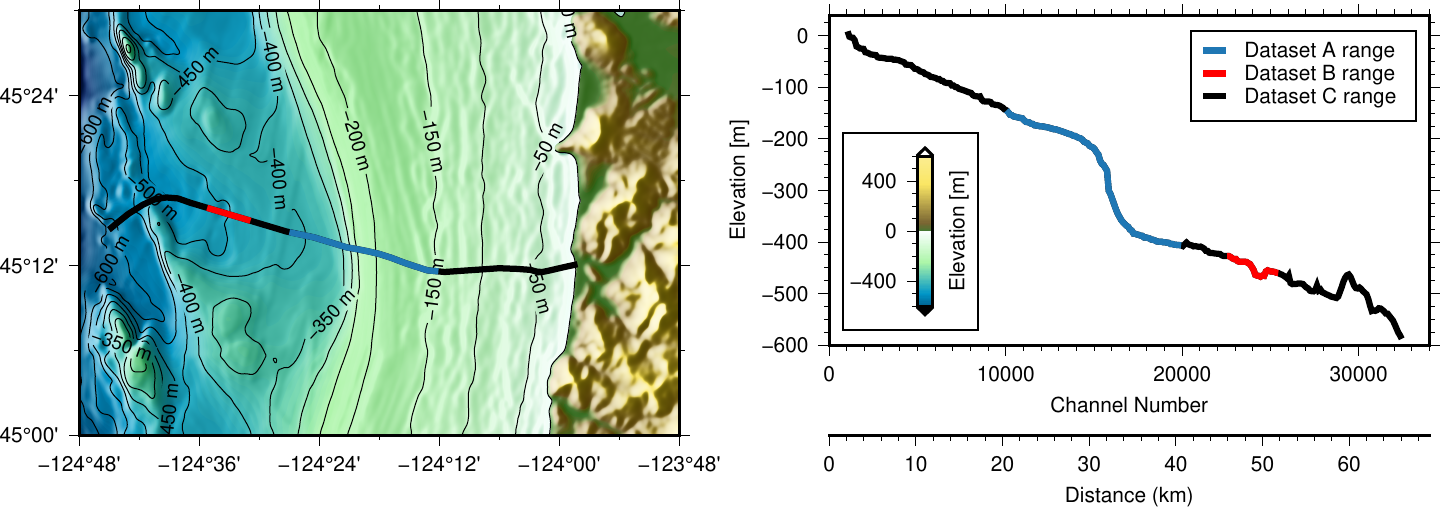}
\caption{(color online) Deployment geometry of the north cable. The fiber portion corresponding to the data analyzed herein is indicated. Dataset C corresponds to the whole cable.}
\label{fig:north_cable_geometry}
\end{figure*}
\begin{table}[t]
\renewcommand{\arraystretch}{1.3}
\caption{\label{tbl:DatasetParameters} List of acquisition parameters and their respective values for each of the datasets used for the analysis.}
\centering
\begin{tabular}{|  p{1.5cm}  |p{1.7cm}  |p{0.8cm}  |p{1.9cm}  |p{1.9cm} |}
%\begin{tabular}{|  p{5cm}  |p{3cm}  |p{0.8cm}  |p{2.5cm}  |p{2.5cm} |}
\hline
\multicolumn{5}{|c|}{Parameter} \\
\hline
description & symbol & unit & \multicolumn{2}{c|}{value} 
\\
\hline
\hline
& & & 
\multicolumn{2}{c|}{Dataset} \\
\cline{4-5}
& & & A, C  
& B \\
\hline
refractive index & $n_{f}$ & - & 1.4682 & 1.4682 \\
\hline
channel spacing & $L_{h}$ & m & 2.042 & 2.042 \\
\hline
number of channels & $n_{h}$ & - & 32600 & 32600 \\
\hline
fiber length & $L_{f} \!= \!L_{h}n_{h}$ & km & 66.57 & 66.57 \\
\hline
pulse width & $T_{p}$ & ns & 250 & 150 \\
\hline
pulse length & $L_{p}\! =\! T_{p}/c_{n}$ & m & 51.08 & 30.65 \\
\hline
gauge length & $L_{g}$ & m & 51.05 (25$L_{h}\!$) & 30.63 (15$L_{h}\!$)\\
\hline
fast-time sampling frequency & $f_{s} = c_{n}/L_{h}$ & GHz & 0.1 & 0.1\\
\hline
slow-time sampling frequency & $f_{r}$ & Hz & 200 & 500  \\
\hline
number of slow-time samples & $n_{t}$ & - & 12000 & 15000 \\
\hline
slow-time duration & $T_{rec}\! =\! n_{t}/f_{r}$ & s & 60 & 30 \\
\hline
\end{tabular}
\end{table}

\section{\label{sec:DataAnalysis} Sound source detection}

The data analysis presented in this section is based on material provided by the Ocean Observatories Initiative (OOI).
The analysis is based on datasets collected by the OOI Regional Cabled Array deployed offshore Central Oregon in November 2021.
This region was covered in the Cascadia initiative, which included 60 Ocean bottom seismometers. \cite{toomey2014cascadia}
The data and associated metadata are publicly available by the University of Washington.\cite{dataOOI}

The analysis herein is based on three datasets.
Datasets A (North-C1-LR-P1kHz-GL50m-Sp2m-FS200Hz\_2021-11-04T022302Z) and B (North-C2-HF-P1kHz-GL30m-Sp2m-FS500Hz\_2021-11-03T015731Z) feature acoustic signals that are attributed to whale calls and ship noise, respectively.\cite{wilcock2023distributed} 
Marine mammal vocalizations and ship noise are loud acoustic sources that are detectable over long distances.
The emanating wavefields insonify a substantial length of the fiber-optic cable and can serve as sources of opportunity for inferring environmental parameters, such as sediment thickness and density.\cite{kuna2021seismic}

Dataset C features lower frequency acoustic signals in the water column due to a distant earthquake. 
Such signals are known as T waves, i.e., Tertiary waves, as they arrive after primary (pressure) P waves and secondary (shear) S waves.\cite{okal2008generation} 
The seismic waves from the earthquake initially propagate through the earth causing high frequencies to be filtered out and then couple into the water column on sloping bathymetry near the earthquake.\cite{degroot2001excitation}
The resulting T waves propagating as acoustic waves in the water column tend to be low frequency with a peak energy below 10 Hz. To capture the full temporal extent of T waves, dataset C concatenates the data in four successive files, namely North-C1-LR-P1kHz-GL50m-Sp2m-FS200Hz\_2021-11-04T093202Z, North-C1-LR-P1kHz-GL50m-Sp2m-FS200Hz\_2021-11-04T093302Z, North-C1-LR-P1kHz-GL50m-Sp2m-FS200Hz\_2021-11-04T093402Z, and North-C1-LR-P1kHz-GL50m-Sp2m-FS200Hz\_2021-11-04T093502Z.

Table~\ref{tbl:DatasetParameters} lists the relevant acquisition parameters for all datasets.
Figure~\ref{fig:north_cable_geometry} shows the deployment geometry of the north cable, from which the datasets above are collected, and indicates the range of the data utilized for the respective analysis.

\begin{figure}[t]
\centering
\includegraphics[width=88mm]{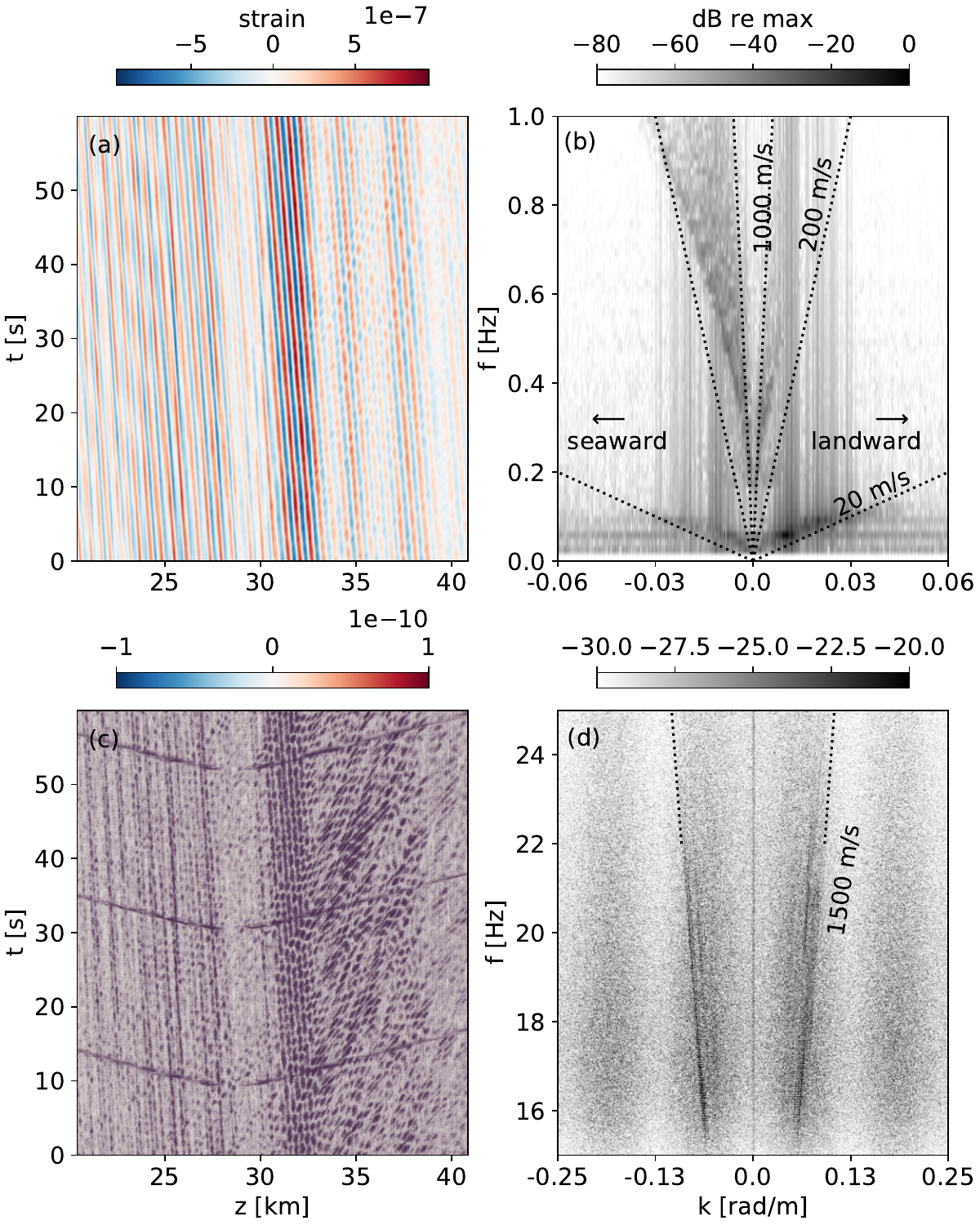}
\caption{(color online) (a) Recorded data in strain units as a function of distance $z$ from the interrogator and slow time $t$ and (b) corresponding f-k representation indicating the dominant very low-frequency energy of ocean and seismic waves. (c) The same data after high-pass filtering featuring acoustic wavefronts propagating along the fiber and (d) the corresponding f-k representation.}
\label{fig:data_filtered_whale_calls}
\end{figure}
\begin{figure}[t]
\centering
\includegraphics[width=88mm]{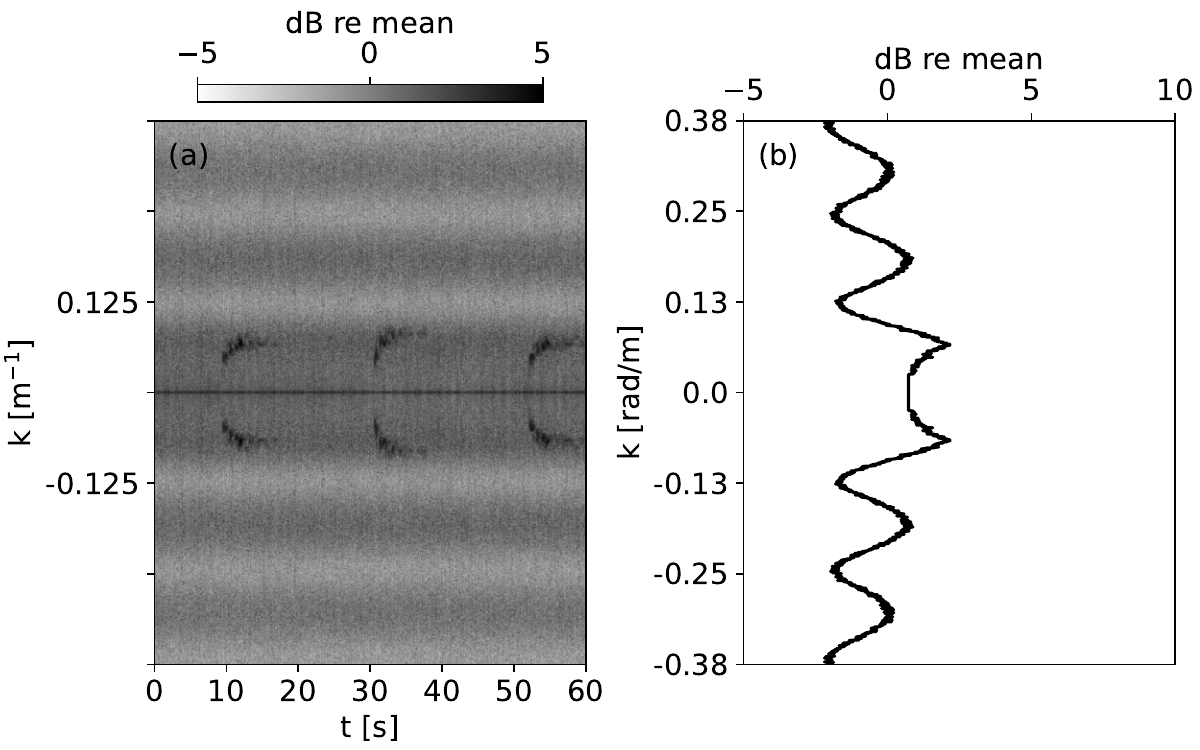}
\caption{(a) Wavenumber-time representation of the data in Fig.~\ref{fig:data_filtered_whale_calls}(c) and (b) the average response across time with notches at multiples of $k=2\pi/L_g=2\pi/51=0.123 {\rm rad/m}$.}
\label{fig:wavenumber_whale_calls}
\end{figure}
\begin{figure}[t]
\centering
\includegraphics[width=88mm]{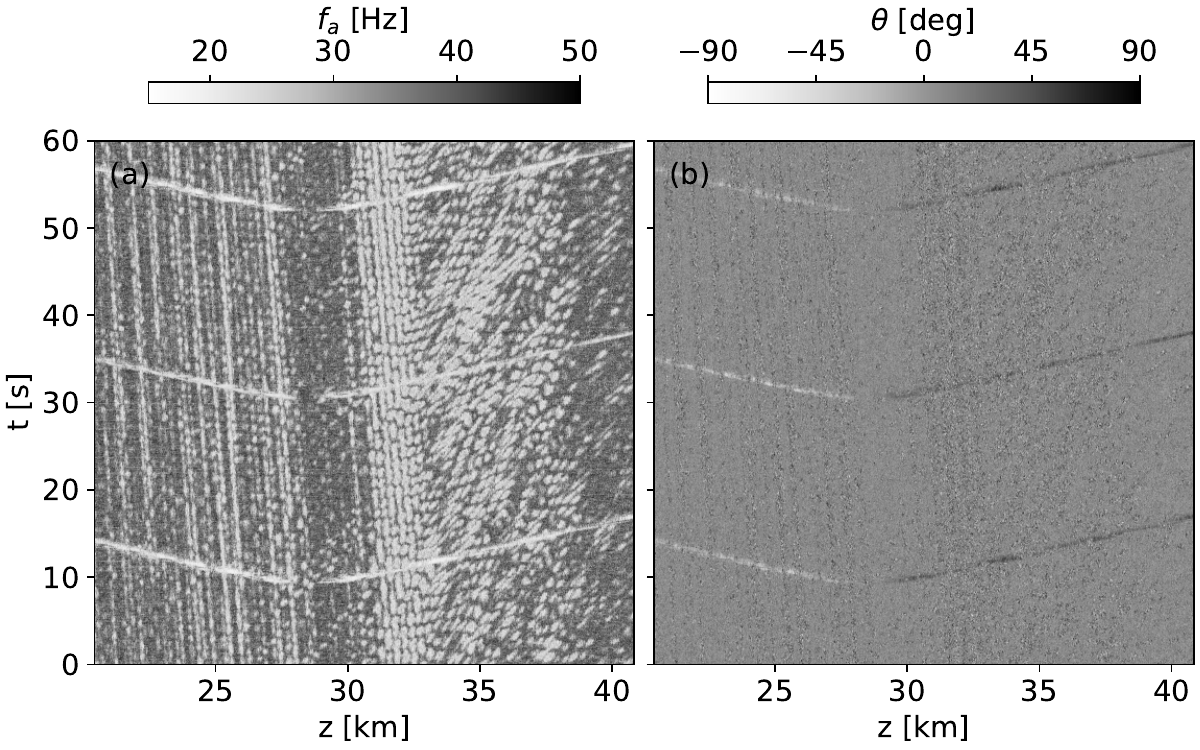}
\caption{Estimates of (a) the frequency $f_{a}$ in Eq.~\eqref{eq:PhaseTemporalGradient} and (b) the angle of incidence $\theta$ 
in Eq.~\eqref{eq:PhaseSpatialGradient} of the acoustic wavefield featuring whale calls sensed by the optical fiber cable.}
\label{fig:hilbert_whale_calls}
\end{figure}
\begin{figure}[t]
\centering
\includegraphics[width=88mm]{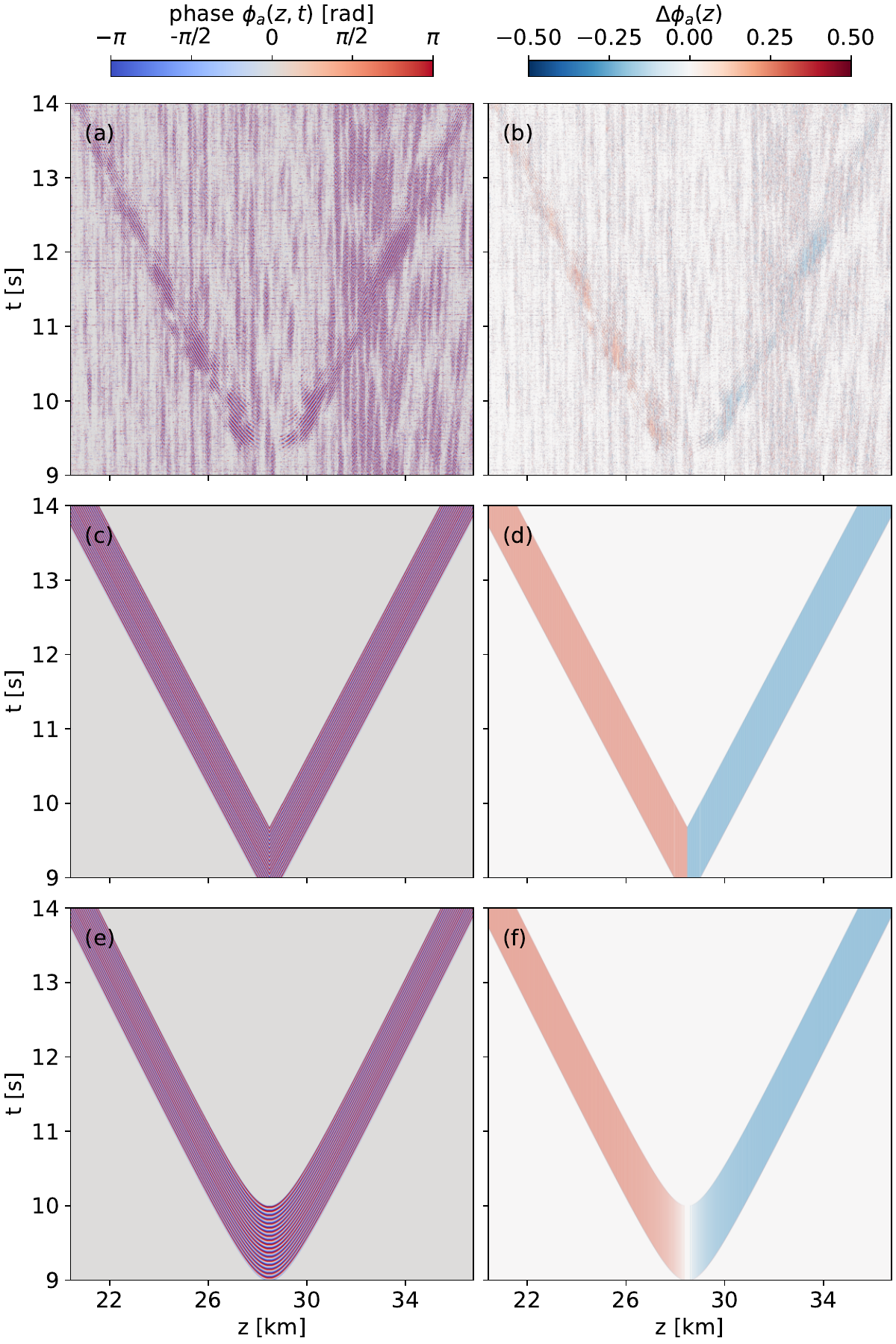}
\caption{(color online) (a) The phase and (b) the spatial phase gradient of the analytical signal in Fig.~\ref{fig:data_filtered_whale_calls}(a) corresponding to the whale call signal at around 10~s in the recording. The corresponding values for a fitted simulation example, assuming either plane wave propagation in (c) and (d)  or spherical wave propagation in (e) and (f).}
\label{fig:phase_whale_calls}
\end{figure}
\begin{figure}[t]
\centering
\includegraphics[width=88mm]{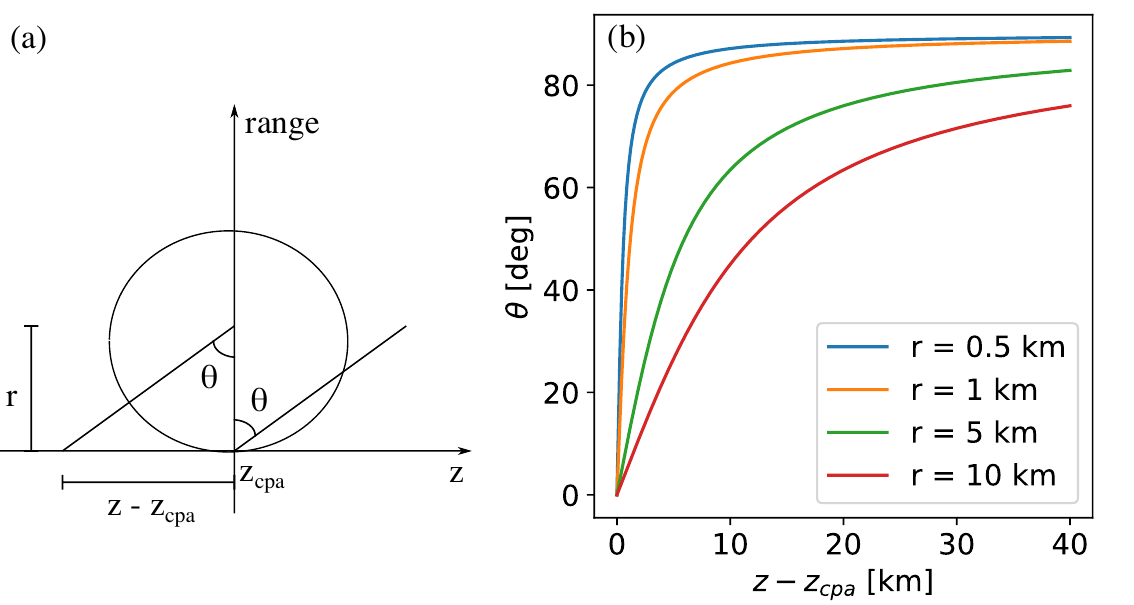}
\caption{(color online) (a) Geometry of spherical wave propagation for an omnidirectional source at range $r$ from the closest point of approach $z_{\rm cpa}$ and (b) corresponding incident angle as a function of axial distance $z$.}
\label{fig:near_field_incidence}
\end{figure}
\begin{figure}[t]
\centering
\includegraphics[width=88mm]{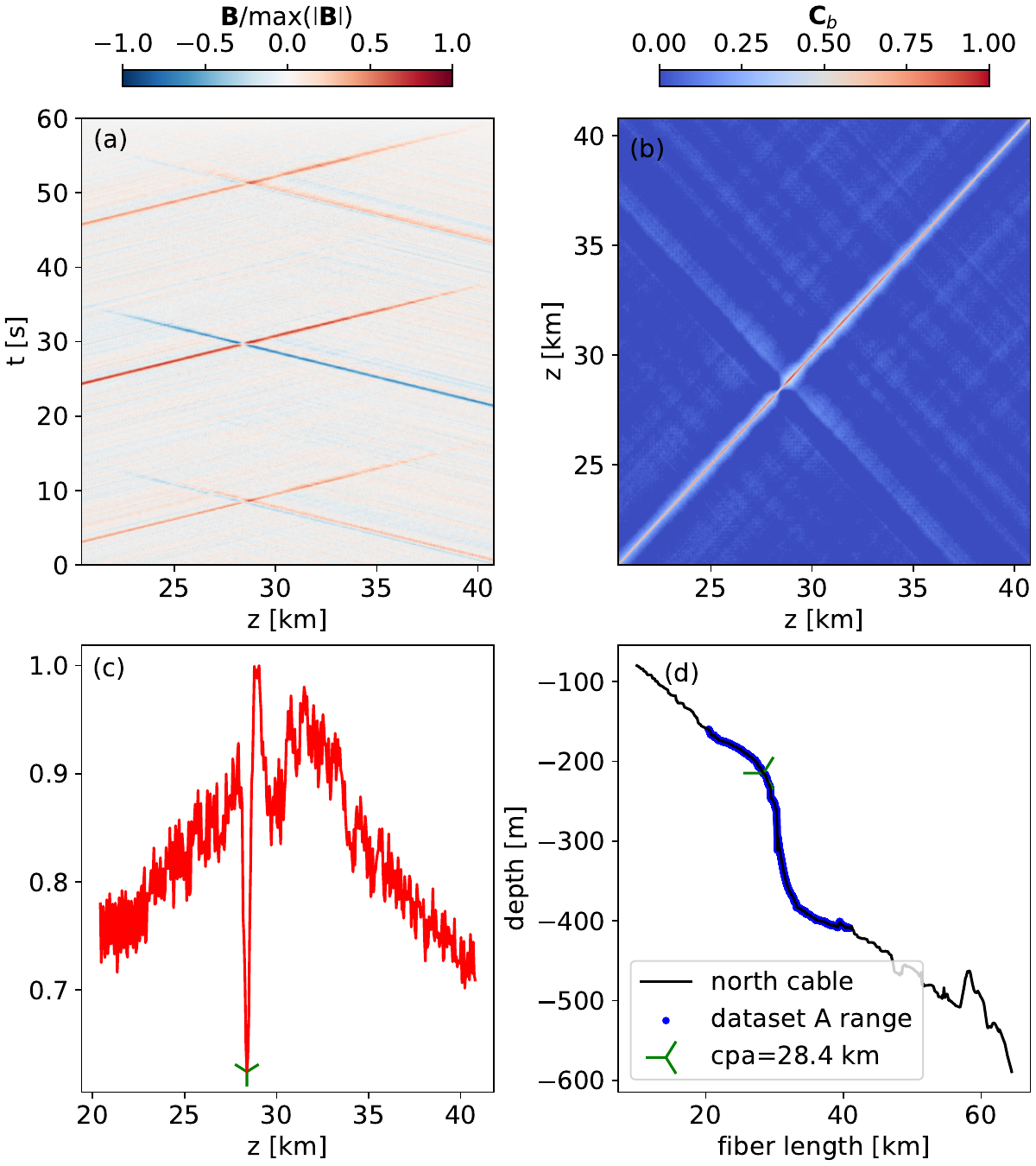}
\caption{(color online) (a) Backpropagation of the spatial phase gradient of dataset A and (b) the corresponding spatial covariance matrix. (c) The closest point of approach is the point with the lowest spatial coherence along the diagonal of the spatial covariance matrix. (d) Location of the closest point of approach on the fiber.}
\label{fig:cpa_whales}
\end{figure}

\subsection{Whale calls}

Dataset A was collected along the transmit fiber in the north cable on November 4, 2021.
The duration of the recording is 60~s sampled at 200 Hz.
The channel spacing is $L_{h}=2.0419$~m and the gauge length is set to $L_{g} = 25L_{h} \approx L_{p} \approx 50$~m.

Figure~\ref{fig:data_filtered_whale_calls}(a) shows the DAS data recorded over 10000 channels (channels 10000--20000) after de-trending, i.e., subtracting the mean per recording and transforming the differential phase measurement to strain, Eq.~\eqref{eq:StrainMeasurement}.
Applying the two-dimensional Fourier transform to the recorded data, the frequency-wavenumber (f-k) representation in Fig.~\ref{fig:data_filtered_whale_calls}(b) is obtained, which shows that the recorded signal is dominated by very low-frequency ocean and seismic waves.\cite{williams2019distributed}
In Fig.~\ref{fig:data_filtered_whale_calls}(b), positive wavenumbers represent wave propagation from the ocean toward the shore and vice versa. 
Signals at the frequency range 0.2--1~Hz are attributed to seismic Rayleigh/Scholte waves with phase velocity in the range 200--1000 m/s and more energy propagating from shore toward the ocean; see Sec.~\ref{sec:OceanBottomSeismology}. 
The lower-frequency signals ($< 0.1$~Hz) represent ocean surface gravity and infragravity waves, with phase velocity ca. 20 m/s and more energy propagating from the ocean toward the shore; see Sec.~\ref{sec:OceanWaves}.

To remove the dominant very low-frequency signals, the data are filtered with a digital eight-order Butterworth high-pass filter\cite{scipy} with a cut-off frequency of 15~Hz.
The high-pass filtered data in time-space and f-k representation, depicted in Figs.~\ref{fig:data_filtered_whale_calls}(c) and (d) respectively, indicate narrowband (15--20 Hz) acoustic arrivals that repeat at regular intervals (every ca. 20~s) and are classified as fin whale (Balaenoptera physalus) calls.\cite{wilcock2023distributed, Goestchel2025fin, watkins1987} 
The f-k spectrum of the filtered data in Fig.~\ref{fig:data_filtered_whale_calls}(d) shows that these signals propagate in both directions along the cable at an apparent phase velocity $c_{a} = 2\pi f/k \approx 1500$~m/s, consistent with an acoustic wave in the water column. 

Figure~\ref{fig:data_filtered_whale_calls}(d) indicates that there are regular notches along the wavenumber dimension. 
To show the wavenumber pattern clearly, Fig.~\ref{fig:wavenumber_whale_calls}(a) shows a representation of the data in Fig.~\ref{fig:data_filtered_whale_calls}(c) in the time-wavenumber domain, by applying the Fourier transform only along the spatial dimension, and Fig.~\ref{fig:wavenumber_whale_calls}(b) shows the average across time.
The notches along the wavenumber dimension occur at multiples of $k = \pm 2\pi/L_{g} \approx \pm 0.125$ rad/m in accordance to the gauge length beampattern; see Fig.~\ref{fig:array_sampling}(d2).

Applying the Hilbert transform\cite{OppenheimBook,scipy} on the real-valued two-dimensional dataset $d(z,t)$ in Fig.~\ref{fig:data_filtered_whale_calls}(c) provides the complex-valued analytic expression for the strain in Eq.~\eqref{eq:NormalStrain}, $\widetilde{\epsilon}_{zz}(z, t) = d(z,t)*\left[\delta(t) + j/\pi t\right]$ such that the instantaneous phase is recovered, i.e.,  
\begin{equation}
\phi_{a}(z, t) = \angle{\widehat{\epsilon}_{zz}(z, t)} = \omega_{a} t - k_{a}z\sin\theta.
\label{eq:AnalyticalSignalPhase}
\end{equation}
The temporal gradient of the unwrapped phase in Eq.~\eqref{eq:AnalyticalSignalPhase} at each location $z_{c}$ is proportional to the acoustic signal frequency as
%
%\begin{equation}
\begin{align}
& \Delta_{t}\phi_{a}(z_{c}, t) = \omega_{a} \Delta t = \frac{2\pi f_{a}}{f_{r}} \quad\Rightarrow \nonumber \\
& f_{a} = \Delta_{t}\phi_{a}(z_{c}, t)f_{r}/(2\pi).
\label{eq:PhaseTemporalGradient}
\end{align}
%\end{equation}
%
Whereas, the spatial gradient of the unwrapped phase in Eq.~\eqref{eq:AnalyticalSignalPhase} at each time instant $t_{c}$ is proportional to the acoustic wavenumber and the incident angle $\theta$ defined in Fig.~\ref{fig:spherical2cartesian} as
%
%\begin{equation}
\begin{align}
&\Delta_{z}\phi_{a}(z, t_{c}) = -k_{a}\Delta z\sin\theta = -\frac{2\pi f_{a}}{c_{a}}L_{h}\sin\theta  \quad\Rightarrow \nonumber \\
&\theta = \arcsin \left(-\frac{\Delta_{z}\phi_{a}(z, t_{c})c_{a}}{2\pi f_{a}L_{h}} \right),
\label{eq:PhaseSpatialGradient}
\end{align}
%\end{equation}
%
where $f_{a}$ is estimated by the temporal gradient of the phase, Eq.~\eqref{eq:PhaseTemporalGradient}.
Figure~\ref{fig:hilbert_whale_calls} shows (a) the acoustic frequency and (b) the incident angle as estimated through the temporal and spatial gradient of the phase of the complex-valued data, respectively.
  
To highlight the characteristics of the V-shaped arrivals attributed to whale vocalizations, Fig.~\ref{fig:phase_whale_calls} details the phase, Eq.~\eqref{eq:AnalyticalSignalPhase}, and the spatial phase gradient, Eq.~\eqref{eq:PhaseSpatialGradient}, of the earliest arrival recorded at around 10~s, and compares the results with a fitted simulation example.
The simulated wave is a sinusoidal pulse of 1~s duration, at a frequency of 18~Hz as estimated from the temporal phase gradient, Eq.~\eqref{eq:PhaseTemporalGradient}.
The simulated pulse is propagating at a sound speed of 1500~m/s and the closest point of approach (CPA) is set at $z_{\rm cpa} = 28.4$~km as inferred from the apex of the wavefront arrival in Fig.~\ref{fig:data_filtered_whale_calls}(c).
Considering plane wave incidence at an angle of 70$^\circ$ as estimated from Eq.~\eqref{eq:PhaseSpatialGradient}, the propagation delay along the $z$ axis is calculated as $t = (z-z_{\rm cpa})\sin\theta/c$.
With the plane wave assumption, the simulation in Figs.~\ref{fig:phase_whale_calls}(c) and (d) is already a good approximation of the recorded data, especially at distances far from the point of closest approach.
Considering spherical, omnidirectional wave propagation from a source at a range of $r=1$~km from the closest point of approach the propagation delay along the $z$ axis is $t = (\sqrt{(z-z_{\rm cpa})^2 + r^2} -r)/c$.
In this case, the simulation in Figs.~\ref{fig:phase_whale_calls}(e) and (f) approximates the data more accurately also at distances around the closest point of approach.
Figure~\ref{fig:near_field_incidence} shows the wavefront angle of incidence, $\theta = \arctan\left((z-z_{\rm cpa})/r\right)$, versus the distance from the closest point of approach $z - z_{\rm cpa}$ for different source ranges $r$. 

The uncertainty in channel locations, the sound speed profile, and the low signal-to-noise ratio make source localization by backpropagation of the amplitude or phase of the recorded signal challenging.
Backpropagating the spatial phase gradient $\Delta_{z}\phi$ of the recorded signal allows an accurate estimation of the location along the cable where the source has its closest point of approach.
Consider a two-dimensional grid in time $t_{0} \equiv t$ and space $z_{0}\equiv z$ with potential spatiotemporal source locations $(t_{0}, z_{0})$, i.e., each point on the grid represents a potential apex of a wavefront as shown in Fig.~\ref{fig:phase_whale_calls}.
Assuming plane wave propagation at $\theta=90^{\circ}$, a reasonable approximation for $z-z_{\rm cpa}>10$~km  as shown in Fig.~\ref{fig:near_field_incidence}, summing the data points over all channels $z$ (that is channels 10000--20000) along the wavefront delays $\left|z-z_{0}\right|/c_{a})$ populates the $(z_{0},t_{0})$ grid as,
\begin{equation}
B(z_{0},t_{0}) =\sum_{z} \Delta_{z}\phi_{a}(z_{0}, t_{0}+\left|z-z_{0}\right|/c_{a}).
\label{eg:Backprop}
\end{equation}
The two-dimensional backpropagation result normalized to the maximum absolute value $\mathbf{B}/\max(\left|\mathbf{B}\right|)$ is shown in Fig.~\ref{fig:cpa_whales}(a).
Since the spatial phase gradient changes sign symmetrically around the wavefront's normal incidence, it will sum up most destructively at the closest point of approach.
Figure~\ref{fig:cpa_whales}(b) shows the spatial covariance matrix $\mathbf{C}_{b} = \mathbf{B}\mathbf{B}^{T}$ of the two-dimensional backpropagation output $\mathbf{B}$ in Fig.~\ref{fig:cpa_whales}(a).
The closest point of approach is identified by locating the minimum of the diagonal of the spatial covariance matrix, $z_{\rm cpa} = \arg\min \left(\text{diag}(\mathbf{C}_{b})\right)$; see Fig.~\ref{fig:cpa_whales}(c).  
\begin{figure}[t]
\centering
\includegraphics[width=88mm]{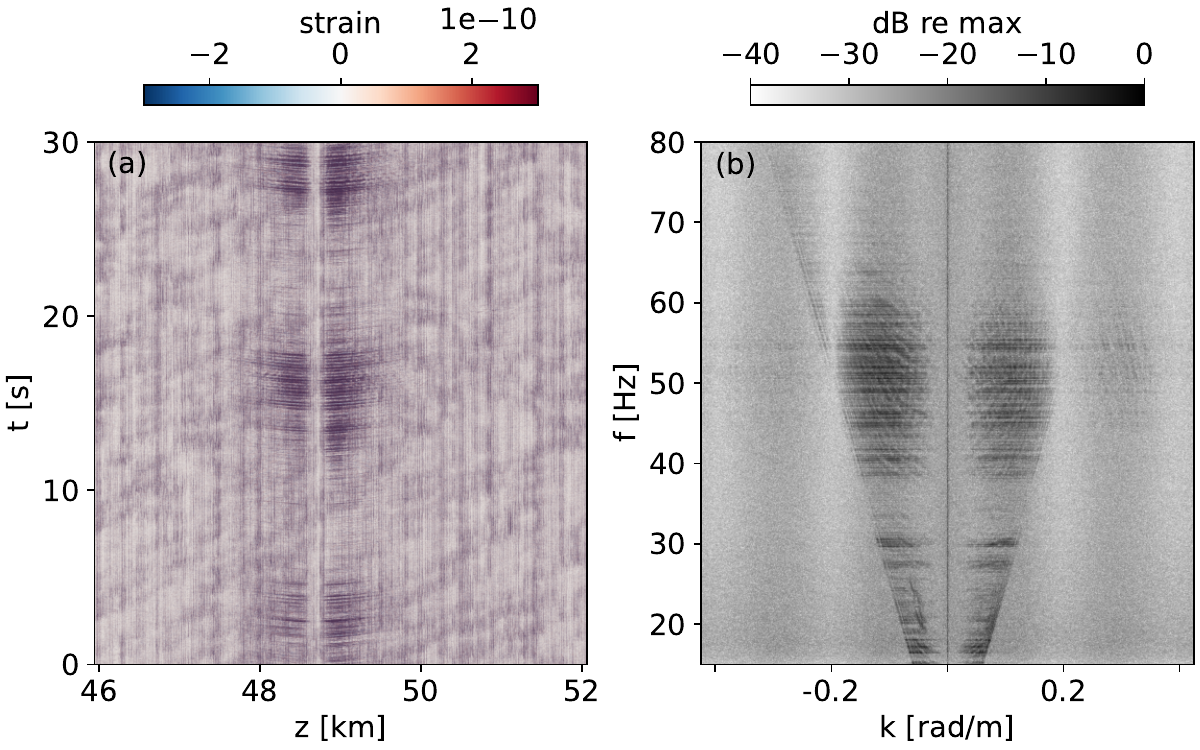}
\caption{(color online) High-pass filtered (cut-off frequency 15~Hz) data featuring a broadband signal attributed to a passing ship. (a) Recordings at successive channels at distance $z$ from the interrogator along the north cable optical fiber. (b) Data representation in the f-k domain.}
\label{fig:data_filtered_ship}
\end{figure}
\begin{figure}[t]
\centering
\includegraphics[width=88mm]{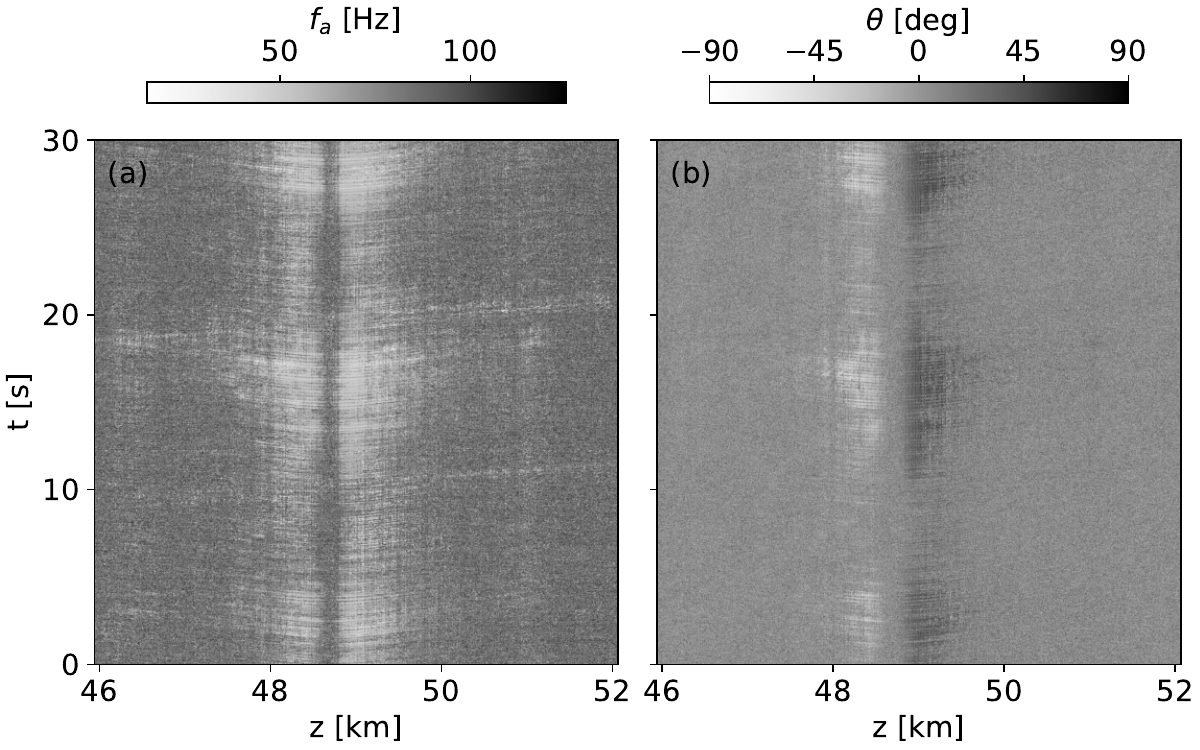}
\caption{Estimates of (a) the frequency $f_{a}$
in Eq.~\eqref{eq:PhaseTemporalGradient} and (b) the angle of incidence $\theta$ 
in Eq.~\eqref{eq:PhaseSpatialGradient} of the acoustic wavefield featuring ship noise sensed by the optical fiber cable.}
\label{fig:hilbert_ship}
\end{figure}
\begin{figure}[t]
\centering
\includegraphics[width=88mm]{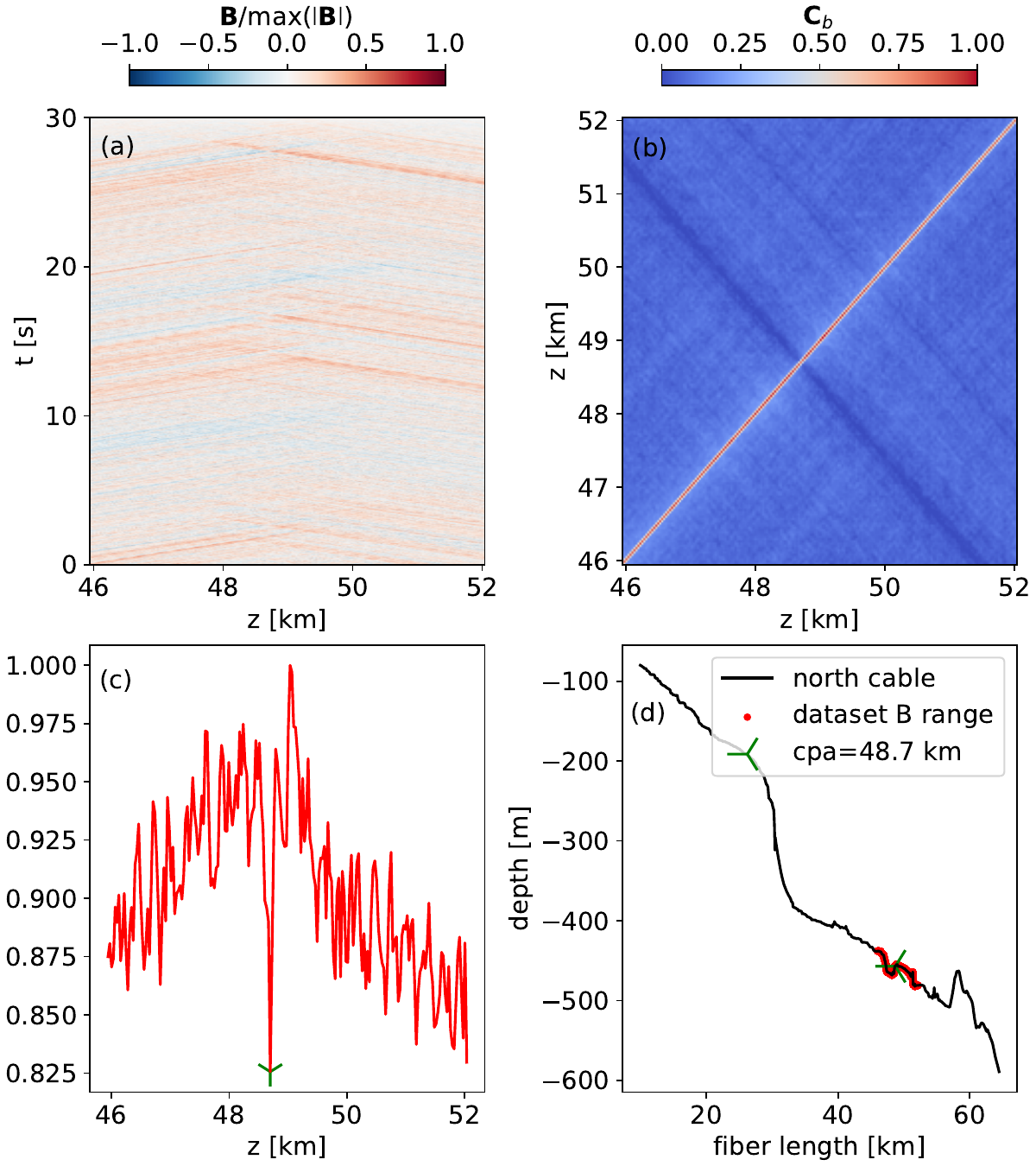}
\caption{(color online) (a) Backpropagation of the spatial phase gradient of dataset B and (b) the corresponding spatial covariance matrix. (c) The closest point of approach has the lowest spatial coherence along the diagonal of the spatial covariance matrix. (d) Location of the closest point of approach on the fiber.}
\label{fig:cpa_ship}
\end{figure}

\subsection{Ship noise}

Dataset B comprises acoustic noise data emitted by a cargo ship, collected along the transmit fiber of the north cable on March 11, 2021. 
The association of the observed signal with ship noise is based on Automatic Identification System (AIS) data\cite{wilcock2023distributed} indicating that the vessel was located at latitude $45.2600^{\circ}$ N and longitude $124.5477^{\circ}$ W at the time of the dataset B recording, traveling at a speed of 12.6 knots.
The duration of the recording is 30~s sampled at 500 Hz. The channel spacing is $L_{h}=2.0419$~m and the gauge length is set to $L_{g} = 15L_{h} \approx L_{p} \approx 30$~m.

Similar to Fig.~\ref{fig:data_filtered_whale_calls}, Fig.~\ref{fig:data_filtered_ship} shows the time-space and f-k representation of the signal recorded over 3000 channels (channels 22500--25500) after de-trending, transforming the measurements to strain units and high-pass filtering at a cut-off frequency of 15~Hz.
The recorded signal in dataset B indicates omnidirectional excitation in the frequency range 40--60 Hz, which is attributed to ship noise.\cite{wilcock2023distributed}
The notches along the wavenumber dimension occur at multiples of $k =  2\pi/L_{g}= 2\pi/30 \approx \pm 0.2$ rad/m.
Figure~\ref{fig:data_filtered_ship}(a) shows that the amplitude of the ship noise oscillates in time with a period of 12--14~s, which corresponds to the average period of the ocean swell during the experiment.
Hence, the observed temporal amplitude modulation can be attributed to changes of the radiated noise due to the ocean swell.\cite{arveson2000radiated} 

Figure~\ref{fig:hilbert_ship} shows (a) the acoustic frequency and (b) the incident angle as estimated through the temporal and spatial gradient of the phase of the complex-valued data, respectively.
Finally, Fig.~\ref{fig:cpa_ship} shows the result of backpropagating the spatial phase gradient $\Delta_{z}\phi$ of the recorded signal and the estimation of the closest point of approach, similar to Fig.~\ref{fig:cpa_whales}.
\begin{figure*}[t]
\centering
\includegraphics[width=150mm]{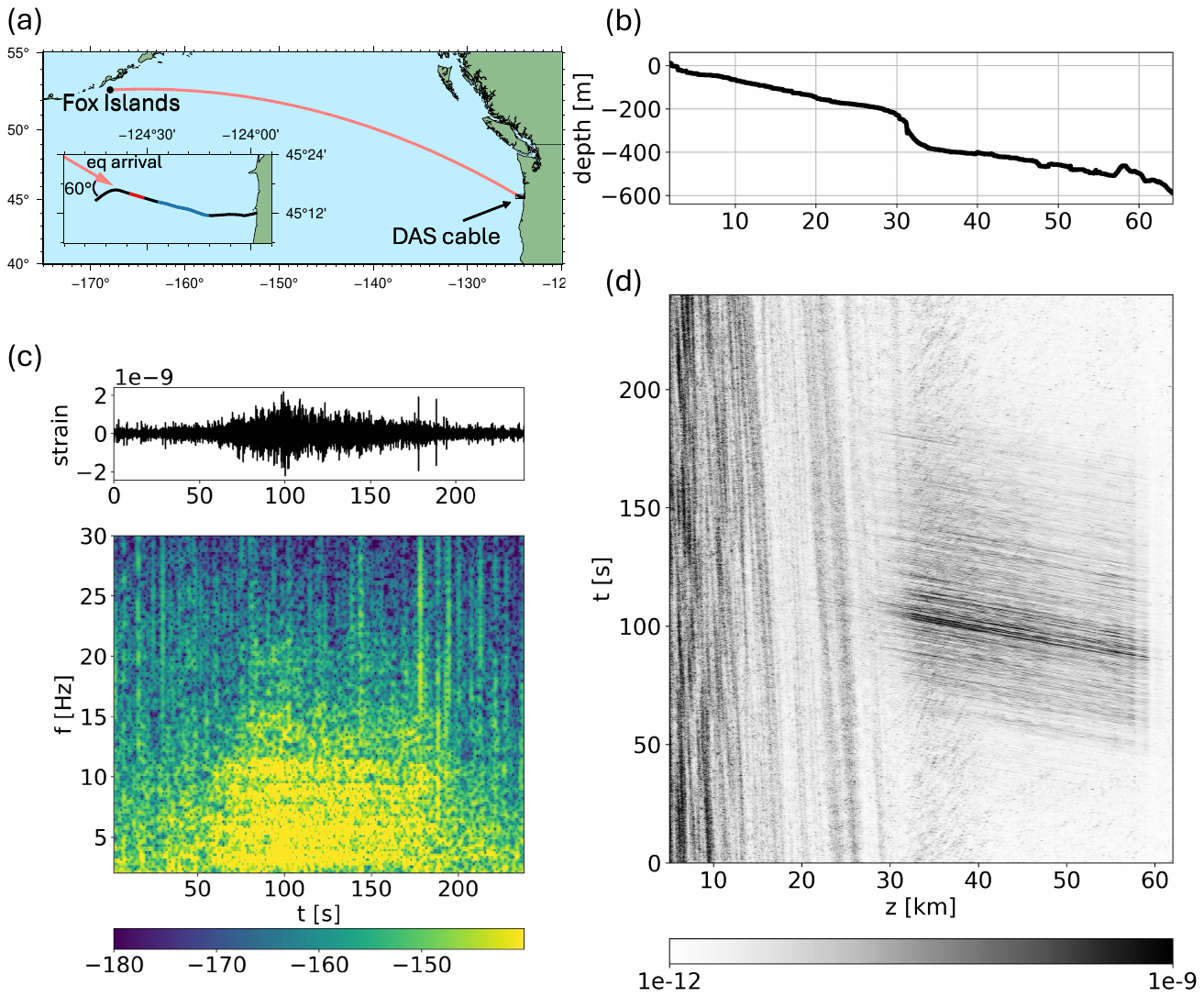}
\caption{(color online)  T wave signal from the Mw 5.2 Fox Islands earthquake on November 4, 2021 at 08:57:06 UTC. (a) Map with great circle arrival of T wave, (b) bathymetry of the north cable deployment, (c) waveform and spectrogram of the recording from a channel at 40 km from the interrogator, and (d) time-space plot of the recordings along the cable.}
\label{fig:t_wave}
\end{figure*}

\subsection{Earthquake noise}

T waves, i.e., acoustic waves coupled into the ocean at a sloping seafloor interface, such as a seamount or the continental margin, due to seismic body waves from nearby earthquakes,\cite{okal2008generation} can be observed with underwater DAS cables at large epicentral distances.\cite{Ugalde2022}
T waves from several earthquakes have been observed in the OOI DAS data. 
Among these, the Mw 5.2 Fox Islands earthquake, which occurred on November 4, 2021, at 08:57:06 UTC, is one of the largest events recorded during this experiment.\cite{shen2024ocean}

Figure \ref{fig:t_wave} illustrates the T wave from this earthquake, observed 35 minutes and 54 seconds after the event on the OOI North Cable.
The data are 1--30~Hz 8th order bandpass Butterworth filtered and then a bandpass f-k filter to retain only the signals with an apparent velocity between 1.4--3~km/s.
An f-k filter processes seismic data by analyzing it in the frequency-wavenumber (f-k) domain.
This technique allows for the isolation or suppression of specific seismic wave components based on their apparent phase velocity.
To apply an f-k filter, the seismic data is first transformed from the time-space domain into the f-k domain.
In the f-k domain, the seismic signal corresponding to the desired apparent velocity is retained. The apparent velocity is the frequency/wavenumber ratio, and in the f-k domain, this is a line with the slope equal to the apparent velocity.
Once the filtering is applied, the data is then transformed back into the time-space domain.

The T wave is observed over a range of around 30~km along the cable, which can be due to a change of the bathymetry reducing the modal cut-off frequency. 
The reduction of the T wave energy at approximately 60 km along the cable is attributed to the directional sensitivity of DAS, explained in Sec.~\ref{sec:AxialStrain}, since the T wave signal arrives at a 120$^\circ$ heading, resulting in a horizontal incident angle of 60--70$^\circ$ with the cable; see Fig.~\ref{fig:t_wave}(a).
 
\section{\label{sec:Discussion}System design considerations}

\subsection{Optical multiplexing}

Most DAS applications have used dark fibers, i.e., redundant network fibers not used for communications, to avoid interference between overlapping DAS pulses with existing optical traffic. 
In subsea networks, most fibers are actively used for telecommunications as the cost and complexity of subsea cable deployment limit the number of installed cables. In contrast, in terrestrial networks dark fibers are more readily available.
Consequently, dark subsea fibers are rare, and the high demand for intercontinental communication further limits their availability. Optical multiplexing methods based on frequency division can overcome this limitation.

Ordinary infrared optical communication typically occurs in the C-band (``Communication"), wavelength 1530--1565 nm, divided into channels. The information to be transmitted is optically encoded by a transceiver operating at one channel's wavelength (or equivalently, frequency), and the output of all transceivers is combined into the same fiber---a technique known as wavelength-division multiplexing (WDM)---to permit simultaneous transmission of independent information. Initially, simultaneous operation of DAS and communication traffic in the same fiber was demonstrated with C-band WDM, dedicating three channels for the DAS.\cite{huang2019first,jia2021experimental,guerrier2022vibration} The disadvantages of this approach are that the communication bandwidth is reduced relative to a fiber without DAS and the peak DAS power must be limited to prevent interference effects.\cite{jia2021experimental} 

Another approach utilizes WDM of C-band communications with DAS operating in the L-band (``Long''), wavelength 1565–-1625 nm, which has only slightly higher attenuation than the C-band but is used much less frequently in communications.\cite{wienecke2023new}
L-band multiplexed DAS has been demonstrated both in laboratory,\cite{wienecke2023new} and in field experiments.\cite{brenne2024non, shi2024multiplexed} 

\subsection{\label{sec:MeasurementRange}Measurement range}

Commercially available DAS systems have a measurement range up to 171~km.\cite{waagaard2021realtime}
Most subsea cables utilize repeaters at regular intervals every 40--100 km to amplify the optical signal and counteract transmission loss.
However, common repeaters are effectively one-way gates preventing any backscattered signal from returning back to the interrogator. 
Consequently, the DAS measurement range is limited to the fiber length that spans the distance between the interrogator and the successive repeater.

To enable multi-span DAS, different optical architectures have been proposed, either utilizing existing repeaters with high-loss loopback (HLLB) paths that return a small fraction of transmitted light via the second fiber of a pair\cite{mazur2024real,ronnekleiv2025range} or bi-directional bespoke repeaters that amplify the backscattered light from distant spans and allow it to propagate backwards.\cite{cunzheng2022246km,ip2022over}
HLLB paths were first used for interferometry\cite{Marra2022,mazur2022transoceanic} and polarization sensing.\cite{costa2023localization}
Multi-span DAS on the HLLB path was demonstrated when combined with optical multiplexing\cite{ronnekleiv2025range} or coherent optical frequency reflectometry.\cite{mazur2024real} 
While these technologies are promising for generalizing DAS across the global subsea cable network, currently none are commercially available and all technologies require significant customization for particular cable configurations.

\subsection{High-frequency signal observation}

While most studies have demonstrated the capabilities of DAS in observing hydro-acoustic signals at frequencies below 100~Hz---such as those produced by earthquakes, baleen whales, and ship noise---the maximum frequency for DAS systems is inherently determined by the pulse repetition rate of DAS recordings, which is limited by the two-way travel time of optical pulses along the cable.
For instance, a 50 km cable allows a pulse repetition frequency of $f_{r} = c_n/(2L_{f})$=2 kHz resulting in a maximum detectable frequency of $f_{\text{max}} = f_{r}/2 =1$~kHz. However, several factors can reduce DAS sensitivity to high-frequency signals:
\begin{enumerate}
\item  Signal Attenuation: Higher-frequency signals attenuate more rapidly in water and in the sediment (for buried cables), leading to weaker signals impinging on the cable.  
Currently, DAS is considerably less sensitive to high-frequency signals than ocean-bottom seismometers or hydrophones. 

\item DAS directivity vs. directional sensitivity: At higher frequencies, the directivity of DAS, which is determined by the gauge length, becomes narrower and the response is severely attenuated by the directional sensitivity of linear DAS cables; see Fig.~\ref{fig:array_sampling_sintheta}. 

\item  Gauge/Pulse length - SNR trade-off: Longer pulses carry more energy, which improves the signal-to-noise ratio (SNR) of the recorded signals. 
However, the longer the pulse width, the larger the gauge length required, since $L_{g}\geq L_{p}$ for unambiguous measurements of the external vibrational field, and consequently the more directive the DAS response resulting in reduced sensitivity to higher frequencies.; see Sec.~\ref{sec:Beampattern}
\end{enumerate}

Despite these challenges, several case studies have demonstrated the potential of DAS for observing acoustic signals above 100 Hz. 
Specifically, ship noise was observed with DAS up to 120 Hz,\cite{Thiem2023ShipNC,cao2023near} whereas a sound source emitting frequencies up to 160 Hz was utilized to localize a DAS cable in shallow water.\cite{shen2024high}
Using a pulse repetition frequency of 2~kHz to interrogate shorter fibers (less than 4~km) with smaller gauge lengths, higher frequency signals from impulsive sources have been recorded with DAS up to 700~Hz\cite{douglass2023distributed}, 960~Hz\cite{taweesintananon2021distributed} and 1~kHz.\cite{Buisman2022highfreqship}
Further reducing the interrogation distance, i.e., for fiber lengths up to 1~km, allows for a higher pulse repetition frequency and smaller gauge lengths (4~m) and, consequently, higher observable frequencies up to 2.5~kHz.\cite{rychen2023test, potter2024first, saw2025distributed}
DAS has also been shown to record signals at frequencies up to 33 kHz.\cite{Buisman2022highfreq}
  
The use of DAS for measuring signals over a broader frequency range remains an active area of research, requiring a deeper understanding of the channel response and DAS sensitivities to high-frequency hydro-acoustic signals.
Further advancements could significantly enhance the applicability of DAS for monitoring underwater soundscapes. 

\subsection{Unknown geometry}

The exact channel locations in DAS are often unknown, unlike hydrophone arrays in ocean acoustics. Particularly, the precise positioning for most telecom cables is either unavailable or proprietary information. Even when the overall cable location is known, estimating the precise channel locations remains challenging. The exact speed of light in optical fiber cannot be measured in situ, leading to uncertainty in converting fast time to distance. Previous studies have demonstrated that a sound source broadcasting near the surface can localize the cable. \cite{shen2024high, Growe2024cableloc}
By extension, ships with AIS location data can serve as opportunistic sources to help determine the cable’s location.

\section{Ocean applications beyond acoustics}

Fiber-optic cables deployed on the seafloor are subject to strain not only due to acoustic waves, but also due to mechanical vibrations from earthquakes, slope movement, ocean waves and bottom currents, as well as being exposed to temperature perturbations.
This section presents DAS applications in seismology and oceanography to aid in the identification of nonacoustic signals in DAS measurements.

\subsection{\label{sec:OceanBottomSeismology} Ocean-bottom seismology}

\begin{figure}[t]
\centering
\includegraphics[width=88mm]{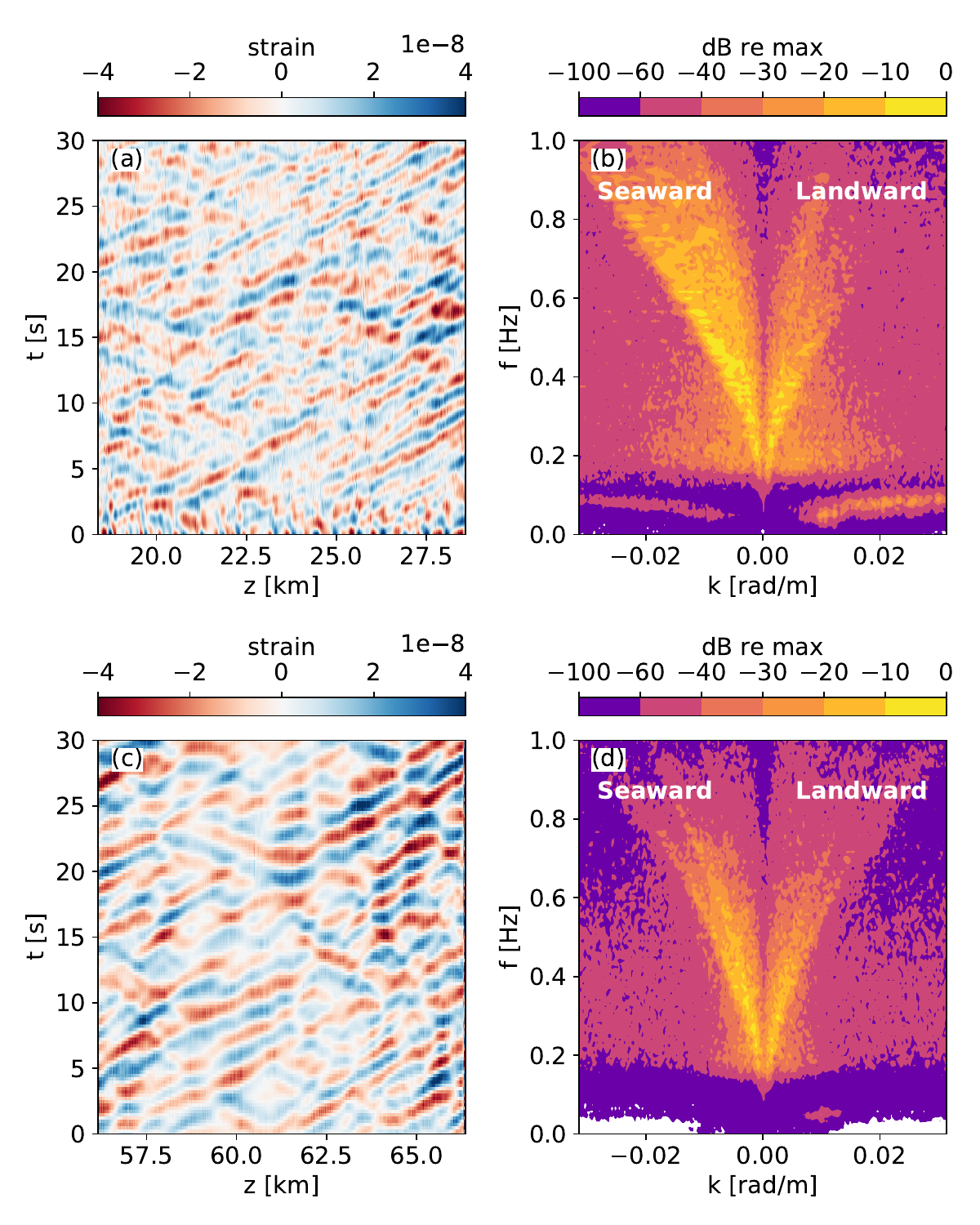}
\caption{(color online) (a) Ambient seismic noise (Scholte waves) from Dataset A along the OOI Regional Cabled Array north cable between channels 9000--14000 at depth $\sim$200 m (see Fig~\ref{fig:north_cable_geometry}), with a second-order Butterworth filter applied between 0.2--1 Hz. (b) Frequency-wavenumber spectrum of the same data segment. Positive wavenumbers indicate propagation towards the shore. (c), (d) Same representations as (a), (b) but for channels 27500--32500 at depth $\sim$500~m.} 
\label{fig:noise}
\end{figure}

Seismic monitoring across the oceans is important for early warning of offshore earthquakes and for improving tomographic models of the deep Earth.\cite{kohler2020plan} 
Fiber-optic sensing on telecommunication networks deployed across the oceans offers a promising solution to increase the global seismographic coverage.
Specifically, DAS has been used for earthquake detection and localization,\cite{lior2021detection,baba2023observation} identification of offshore fault zones from scattered waves,\cite{lindsey2019illuminating} and quantification of linear and nonlinear earthquake ground motion.\cite{viens2022nonlinear}
Recent tests suggest that seismic monitoring with DAS on submarine cables can reduce the early warning time (up to 10 s) for offshore earthquakes.\cite{romanowicz2023seafoam,yin2023real}

The horizontal orientation of fiber-optic cables on the seafloor makes DAS particularly sensitive to shear and Scholte (water-sediment interface) waves.\cite{martin2021introduction,spica2020marine}
Shear-wave resonances observed with DAS can be used to infer the thickness of submarine sediments,\cite{trabattoni2024sediment,taweesintananon2024near} while the frequency-dependent dispersion characteristics of surface waves have been used to infer the depth-dependent shear wave velocity profile.\cite{spica2020marine,cheng2021utilizing,williams2021scholte}

Ocean-bottom DAS observations of seismic noise is seen in the 0.2--1 Hz noise radiating from the coast in f-k spectra from the OOI Regional Cabled Array (Fig.~\ref{fig:noise}). 
Others have used DAS to characterize microseisms.\cite{williams2019distributed,guerin2022quantifying,xiao2022locating,fang2022unraveling}

\subsection{\label{sec:OceanWaves} Ocean surface waves, nearshore processes, and tsunamis}

\begin{figure}[t]
\centering
\includegraphics[width=88mm]{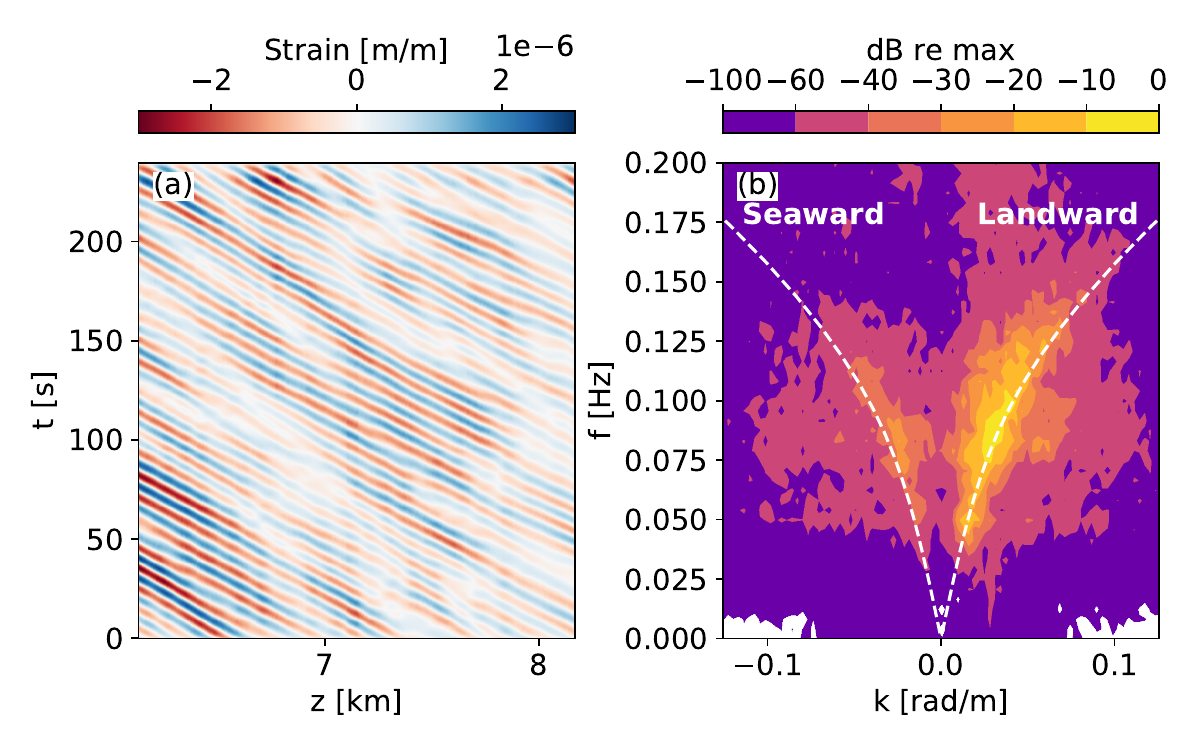}
\caption{(color online) (a) Ocean surface gravity waves observed between channels 3000--4000 (depth $\sim$50 m) on the OOI Regional Cabled Array north cable, band-pass filtered 0.05--0.5 Hz. The data are concatenated from four successive files starting from the file of Dataset A. (b) Frequency-wavenumber spectrum of the same data segment, showing propagation towards the shore at positive wavenumbers and the theoretical dispersion relation of surface gravity waves (white dotted line).} 
\label{fig:sgw}
\end{figure}

Very low frequency ($<$0.5~Hz) pressure perturbations from ocean surface gravity waves (SGW) also induce strain in a fiber-optic cable deployed on the seafloor, which can be measured with DAS \cite{lindsey2019illuminating,sladen2019distributed,williams2019distributed} and used to quantify sea surface characteristics, such as significant wave height and average wave period. \cite{glover2024comparisons,glover2024measuring,meule2024reconstruction,smith2023observations}
Figure~\ref{fig:sgw} shows DAS observations of SGW along the north cable of the OOI Regional Cabled Array and their dispersion characteristics in the f-k spectra.

The dispersion relation for SGW [shown with a dotted line in Fig.~\ref{fig:sgw}(b)] varies with water depth $h$ as 
\begin{equation}
\omega^2 = g k \tanh(kh),
\label{eq:Dispersion}
\end{equation}
where $\omega = 2\pi f$ is angular frequency at frequency $f$, $k=2\pi/\lambda$ is wavenumber at wavelength $\lambda$, and $g$ is gravitational acceleration.
In deep water, i.e., for $h\ge \lambda/(2\pi)$, $\tanh(kh)\approx 1$ and $\lambda = g/(2\pi f^{2})$ from Eq.~\eqref{eq:Dispersion}, hence wind waves and swell with frequencies around 0.05--0.5~Hz have wavelengths around 600--6~m.
In shallow water, i.e., for $h\ll\lambda/(2\pi)$, $\tanh\left(kh \right)\approx kh$ and $\lambda = \sqrt{gh}/ f$, hence the wavelength scales as $\sqrt{h}$, asymptotically approaching zero at the beach.
The choice of an appropriately short gauge length and pulse width (see Eq.~\eqref{eq:NormalStrain}) is, therefore, essential to resolve the smaller wavelengths in shallow water and accurately estimate parameters like significant wave height.\cite{glover2024comparisons}

For a plane harmonic SGW with amplitude $A$ at the sea surface, the ratio of the pressure at the seafloor $p$ to the pressure at the sea surface $p_0 = \rho g A$ is \cite{bromirski2002near}
\begin{equation}
    \frac{p}{p_0} = \frac{1}{\cosh(k h)},
    \label{eq:Pressure}
\end{equation}
where $k$ follows the dispersion relation in Eq.~\eqref{eq:Dispersion}. 
In shallow water ($h\ll\lambda/(2\pi)$), $\cosh(kh)\approx 1$, hence the pressure $p=p_{0}$ does not depend on water depth.
Conversely, in deep water ($h\geq \lambda/(2\pi)$), $\cosh(kh)\approx e^{kh}/2$, hence the pressure $p=p_{0}e^{-kh}/2$ decays exponentially with depth.
For example, at 1~m water depth, a 0.02~Hz and a 0.2~Hz wave exert the same pressure at the seafloor, whereas, at 100~m water depth, the pressure due to a 0.02~Hz wave is more than $10^7$ times larger than the pressure due to a 0.2~Hz wave.\cite{taweesintananon2023distributed}

Consequently, water depth acts as a low-pass filter for the SGW spectrum, so DAS can only observe wind waves and swell ($\sim$0.05--0.5~Hz) in shallow water near the coast where the seafloor pressure is finite. 
Infragravity waves ($\sim$ 0.003--0.03~Hz) and tsunamis ($\lesssim 0.02$~Hz) are generally observable at all depths.\cite{bromirski2002near,taweesintananon2023distributed,glover2024measuring,shi2024multiplexed} 

DAS enables wave measurement in a continuous transect from the beach to the outer shelf, and, since cables are often buried in shallow water, DAS is generally immune to strong currents or sea ice. 
DAS has been used to measure the attenuation of SGW in sea ice\cite{smith2023observations} and track tsunamis generated by glacial calving events.\cite{graeff2024shaken}
Array processing of DAS data has been used to assess the influence of coastal reflection of SGWs on microseism generation\cite{guerin2022quantifying} and quantify wave-current interaction in shallow water.\cite{williams2019distributed,williams2022surface,lin2024monitoring} 
Observation of infragravity waves and small tsunamis with DAS\cite{shi2024multiplexed,tonegawa2024high,xiao2024detection} suggests that the existing global subsea network of fiber-optic cables can be used for real-time tsunami detection and early warning.

\subsection{Temperature sensing}

In addition to mechanical strain, DAS can directly measure temperature changes.
For a temperature change $\Delta T$, the associated phase change over a gauge length $L_g$ is a function of changes in the optical path length ($\Delta L/L_g$) and refractive index ($\Delta n/n$), similarly to Eq.~\eqref{eq:DifferentialPhasevsStrain} [\onlinecite[Eq.~(16)]{hocker1979fiber}],
\begin{align}
\Delta \Phi &= \frac{4\pi n L_g}{\lambda_{p}} \left( \frac{\Delta L}{L_g} + \frac{\Delta n}{n}\right) \nonumber \\
&\approx \frac{4\pi n L_g}{\lambda_{p}} \left( \alpha_T + \psi \right) \Delta T,
\label{eq:DAStemperature}
\end{align}
where $\alpha_T = L_g^{-1} \partial L/\partial T$ represents the thermo-elastic effect (expansion of the fiber with increasing temperature) and $\psi = n^{-1} \partial n/\partial T$ represents the thermo-optic effect (slower propagation of light with increasing temperature).\cite{hocker1979fiber,sidenko2022experimental} 
Substituting typical values of these coefficients for silica fiber,\cite{hocker1979fiber} the sensitivity to temperature is $\Delta\Phi/\Phi = 7\times10^{-6}$ per 1 $^\circ$C temperature change, where $\Phi = 4\pi n L_g/\lambda_{p}$. 
Comparing this with the strain sensitivity in Eq.~\eqref{eq:DifferentialPhasevsStrain}, a temperature change of 1$^\circ$C induces the same phase change $\Delta \Phi $ as a mechanical strain of $10^{-5}$ m/m.\cite{zumberge2018measuring}

% Use this paragraph for comparison of methods
Equation~\eqref{eq:DAStemperature} indicates that DAS can be also used for distributed measurements of temperature changes.
More common temperature sensing methods, i.e, distributed temperature sensing (DTS) based on Raman scattering and distributed strain and temperature sensing (DSTS) based on Brillouin scattering (Fig.~\ref{fig:scattering_mechanisms}),
measure the amplitude rather than the phase of backscattered light.\cite{Hartog2017}
Consequently, the sensitivity of both DTS and DSTS systems decreases with the measurement range, which can only be partly mitigated by increasing the temporal or spatial averaging.\cite{selker2006distributed}
One advantage of DAS is its long measurement range;\cite{koyamada2009fiber,williams2023fiber} see Sec.~\ref{sec:MeasurementRange}.

DAS is more sensitive to very small (milliKelvin-scale) temperature fluctuations than DTS or DSTS.\cite{koyamada2009fiber,seabrook2022comparison}
However, DAS can only measure \emph{relative} temperature, whereas DTS can be calibrated to measure the \emph{absolute} temperature.\cite{sidenko2022experimental,sinnett2020distributed}
DTS only measures temperature and is unaffected by strain. In principle, DSTS can independently measure both strain and temperature by combining information from the Brillouin frequency shift and scattering power.\cite{clement2021b}
By contrast, temperature and strain contributions in DAS measurements cannot be disentangled, unless probing simultaneously multiple fibers with different thermal properties.\cite{zumberge2018measuring}
Cable burial insulates DAS from temperature transients, and temperature effects can generally be ignored at typical seismic and acoustic frequencies even on unburied cables.\cite{williams2023fiber} 

% Use this paragraph for condensed references
In geophysics, DAS has been used for distributed dynamic temperature sensing to monitor fluid flow in boreholes.\cite{sidenko2022experimental,titovspe,karrenbach2019fiber,bradley2024}
Offshore DAS recordings in Japan\cite{ide2021very}, the Canary Islands\cite{williams2023fiber}, and the Mediterranean Sea\cite{pelaez2023high} have revealed low-frequency transients over timescales from minutes to weeks, attributed to temperature oscillations from internal waves and turbulence.
These are qualitatively consistent with observations from Raman-based DTS deployed on seafloor cables\cite{sinnett2020distributed,davis2020fate} and quantitatively consistent with collocated DTS\cite{graeff2024shaken} and DSTS measurements\cite{pelaez2023high} in two cases.

\subsection{Flow-related strain}

Strain measurements from fiber-optic cables that are in loose mechanical contact with the surrounding medium, such as freely suspended cables in the water column or cables lying on the seafloor, are subject to flow-induced cable vibrations, which may mask or distort acoustic signals. \cite{Ugalde2022,mata2023identification}
The flow-related vibrations are called coupling noise in borehole applications, where the DAS cable is clamped to the wall of a borehole.\cite{daley2016field}
In an ocean setting, such vibrations might offer a unique insight into hydrodynamics when driven by near-bottom flows.\cite{mata2023monitoring,spingys2024distributed,spingys2024optical} 

Fluid mechanics predict that flow past a cable creates vortices at either side of the cable. 
The pressure from each vortex drives harmonic oscillations of the cable at a frequency that is given by the Strouhal number, $\text{St} = f D/v$, where $f$ is the vibrational frequency, $D$ is the diameter of the cable, and $v$ is the speed of the flow across the cable.\cite{williamson2004vortex} 
For a typical Strouhal number of 0.2 and a telecommunication cable diameter of 2 cm, flow speeds from 0.01--1 m/s induce vibrations at the frequency range $f=v\text{St} /D\approx$ 0.1--10 Hz on free-hanging cable segments, which can pose a challenge for seismo-acoustic studies in that frequency range.

In theory, the flow speed in the plane orthogonal to the cable can be recovered from DAS data.
A pioneering study\cite{mata2023monitoring} compared the vibrational frequencies of a free-hanging DAS cable segment with a collocated flow velocity measurement, demonstrating the validity of this approach.
However, the dynamics of the vibrational modes are affected by the cable's length and stiffness making flow-induced strain measurements with DAS challenging.\cite{mata2023identification}
Flow-induced vibrations were later exploited to infer the particle velocity of passing internal waves in a Greenland fjord.\cite{graeff2024shaken}

\section{\label{sec:Conclusions}Conclusions}

This review provides an encompassing description of DAS technology and demonstrates its potential applications to ocean acoustics.
We detail how DAS converts a fiber-optic cable into a distributed sensor of vibrational fields, such as propagating sound, substantiating that active optic sensing can be used as a proxy for passive acoustic monitoring of the environment.
Based on the mathematical derivation of the underlying physics from electromagnetic to mechanical and acoustic quantities in DAS measurements for coherent systems, we discuss the effect of typical DAS acquisition parameters in signal processing.
The potential of DAS technology for underwater acoustic applications, such as sound source detection, is demonstrated with the Ocean Observatories Initiative data.

Deriving the directional sensitivity of the distributed sensing modality in terms of conventional array signal processing, we explain that a limitation of current DAS systems in observing high-frequency signals stems from competing data acquisition requirements.
Specifically, the maximum detectable frequency is determined by the repetition frequency of the pulse that interrogates the optical fiber, which, in turn, is inversely proportional to the two-way travel time of the pulse along the fiber segment that defines the measurement range.
Additionally, the longer the transmitted pulse length, the higher the signal-to-noise ratio of the measurand but the less sensitive the system becomes to high-frequency vibrational fields.

The extensive coverage and robustness to harsh environmental conditions make fiber-optic sensing a considerable alternative to conventional discrete arrays of sensors.
This review provides the background and insight to expand the application of DAS in monitoring the oceans.

%  Final sections -------------------------------------------------------- %
\section{Author Declarations}
The authors have no conflict of interest to disclose.

\section{Data Availability}
The data that support the findings of this study are publicly available by the University of Washington.\cite{dataOOI}

\begin{acknowledgments}
This work was performed while Angeliki Xenaki was employed at the NATO-STO Centre for Maritime Research and Experimentation and Peter Gerstoft was a visiting scientist.
The work was funded by the NATO Allied Command Transformation and by the Office of Naval Research under Grant N00014-21-1-2267.
The data supporting this study were collected by the Ocean Observatories Initiative (OOI), a major facility fully funded by the National Science Foundation under Cooperative Agreement No. 1743430, and the Woods Hole Oceanographic Institution OOI Program Office.
\end{acknowledgments}

\appendix*
\section{\label{sec:COTDR}Coherent optical reflectometry}

Optical time-domain reflectrometry detects the backscattered light from a pulse transmitted in an optical fiber with a photodetector in the interrogator [\onlinecite[Sec. 3.1]{Hartog2017}].
The photodetector's output current $I_{pc}$ is proportional to the power of the received electric field, $I_{pc}\propto \lvert E_{b} \rvert^{2}$, hence the phase information of the backscattered signal is discarded.
To retain the phase information of the backscttered light, coherent optical reflectrometry methods are used.

In principle, coherent optical reflectometry mixes the backscattered signal returning from the interrogated fiber $E_{b}$ with a reference signal $E_{\text{ref}}$.
The resulting signal is fed to the photodetector such that the output current $I_{pc}$ is proportional to the power of the resulting interferometric field,\cite{He2021}
\begin{equation}
\begin{aligned}
I_{pc}  & \propto \Re\left\{ \left(E_{\text{ref}} + E_{b} \right)\left(E_{\text{ref}}+  E_{b} \right)^{*} \right\}\\
&= \lvert E_{\text{ref}} \rvert^{2} + \lvert E_{b} \rvert^{2} + 2\Re\left\{E_{\text{ref}}E^{*}_{b}\right\}.
\end{aligned}
\label{eq:CurrentPhotodetector}
\end{equation}

The reference signal $E_{\text{ref}}$ is, usually, a replica of the transmitted pulse $E_{p}$, potentially amplified and frequency shifted by an optical local oscillator,
\begin{equation}
\widetilde{E}_{\text{ref}}(\tau) = G_{\text{LO}}e^{j\Delta\omega \tau }\widetilde{E}_{p}(\tau),
\label{eq:LOsignal}
\end{equation}
where $G_{\text{LO}}$ is the gain and $\Delta\omega$ the frequency shift introduced by the local oscillator.
Deriving from Eqs.~\eqref{eq:LOsignal},~\eqref{eq:ProbePulseComplex}, and~\eqref{eq:BackscatterTotalComplex}, the interference product becomes,
\begin{equation}
\begin{aligned}
&\widetilde{E}_{\text{ref}}(\tau)\widetilde{E}^{*}_{b}(\tau)   \\
&=\quad G_{\text{LO}}e^{j\Delta\omega \tau }\widetilde{E}_{p}(\tau)\Big( \int\limits_{z-L_{p}/2}^{z+L_{p}/2} \!\! \widetilde{E}_{p}\left(\tau \!-\! 2\frac{z_{s}}{c_{n}}\right)s(z_{s})\mathrm{d}z_{s} \Big)^{*} \\
&=G_{\text{LO}}e^{j\Delta\omega \tau }p(\frac{\tau}{T_{p}})e^{j\omega_{p}\tau}\\
& \quad\cdot \Big(\int\limits_{z-L_{p}/2}^{z+L_{p}/2} \!\! w(z-z_{s})\! e^{j\omega_{p}(\tau - 2\frac{z_{s}}{c_{n}})}s(z_{s})\mathrm{d}z_{s} \Big)^{*}\\
&=G_{\text{LO}}e^{j\left(\Delta\omega \tau + \phi(\tau)\right)} \int\limits_{z-L_{p}/2}^{z+L_{p}/2} \!\! w(z)w^{*}(z-z_{s})\! e^{j2k_{p}z_{s}}s^{*}(z_{s})\mathrm{d}z_{s}\\
& =G_{\text{LO}}e^{j\left(\Delta\omega \tau + \phi(\tau)\right)} \widetilde{q}(z), 
\end{aligned}
\label{eq:COTDRsignal}
\end{equation}
where $\widetilde{q}(z) = \int_{z-L_{p}/2}^{z+L_{p}/2}  w(z)w^{*}(z-z_{s}) e^{j2k_{p}z_{s}}s(z_{s}^{*})\mathrm{d}z_{s} = q(z)e^{j\phi(z)}$ is introduced for notational brevity and $\phi(z)$ accounts for the phase introduced by changes in the optical path between the reference and the backscattered signal from location $z$ along the fiber received at time $\tau$.

Hence, the photodetector output in Eq.~\eqref{eq:CurrentPhotodetector} comprises the contributions of the signals $\widetilde{E}_{\text{ref}}$ at frequency $\omega_{p} + \Delta\omega$ [Eq.~\eqref{eq:LOsignal}], $\widetilde{E}_{b}$ at frequency $\omega_{p}$ [Eq.~\eqref{eq:BackscatterTotalComplex}], and the interference signal $\widetilde{E}_{\text{ref}}\widetilde{E}^{*}_{b}$ at the difference frequency $\Delta\omega$ [Eq.~\eqref{eq:COTDRsignal}].
Lowpass filtering the photodetector output $I_{pc}$ in Eq.~\eqref{eq:CurrentPhotodetector} excludes the high-frequency contribution of the signals $|\widetilde{E}_{\text{ref}}|^2$ and $|\widetilde{E}_{b}|^2$ and retains the interference signal at the difference frequency $\Delta\omega$
\begin{equation}
\begin{aligned}
I_{\text{lp}}(\tau)  &\propto \Re\left\{\widetilde{E}_{\text{ref}}(\tau)\widetilde{E}^{*}_{b}(\tau)\right\} \\
& =\quad G_{\text{LO}}q(z)\cos \left(\Delta\omega \tau \!+\! \phi(z)\right)~.
\end{aligned}
\label{eq:CurrentPhotodetectorBaseband}
\end{equation}

Couplers can extract the phase of the detected signal [\onlinecite[Sec. 3.3.1.1]{Hartog2017}].
For example, an in-phase/quadrature (I/Q) demodulation system\cite{wang2016coherent} uses a $2\times 2$ channel coupler to output both the in-phase signal, Eq.~\eqref{eq:CurrentPhotodetectorBaseband}, and the corresponding quadrature signal by phase shifting the in-phase signal by $\pi/2$,
\begin{equation}
Q_{\text{lp}}(\tau) \!=\! I_{\text{lp}}(\tau, \pi/2)  \!\propto \! G_{\text{LO}}q(z)\sin \left(\Delta\omega \tau + \phi(z)\right).
\label{eq:QuadratureCurrentPhotodetectorBaseband}
\end{equation}
Then, the phase of the detected signal is calculated as
\begin{equation}
\phi(z) = \arctan\left( \frac{Q_{\text{lp}}(\tau)}{I_{\text{lp}}(\tau)} \right) - \Delta\omega \tau.
\label{eq:IQdemodulationPhase}
\end{equation}
By measuring the differential phase between the mixed signals, coherent optical reflectometry provides a measurand that is linearly related to the distance traveled by the backscattered wave.

%  Bibliography --------------------------------------------------------- %
%\bibliography{bibliography/das}
%\bibliography{das_cleaned_corrected}

\end{document}